\documentclass[]{emulateapj}
\usepackage{longtable}
\usepackage{graphics}
\usepackage{rotating}
\usepackage{amsmath}
\usepackage{natbib}
\usepackage{hyperref}
\usepackage{color,soul}
\def\gtorder{\mathrel{\raise.3ex\hbox{$>$}\mkern-14mu
             \lower0.6ex\hbox{$\sim$}}}
\def\ltorder{\mathrel{\raise.3ex\hbox{$<$}\mkern-14mu
             \lower0.6ex\hbox{$\sim$}}}

\slugcomment{\today}
\shorttitle{Radio emission from PTF broad-lined Type Ic supernovae}

\shortauthors{Corsi et al.}

\begin{document}

\title{Radio observations of a sample of broad-lined type Ic supernovae discovered by PTF/iPTF: A search for relativistic explosions}
\author{A.~Corsi\altaffilmark{1}, A.~Gal-Yam\altaffilmark{2}, S.~R.~Kulkarni\altaffilmark{3}, D.~A.~Frail\altaffilmark{4}, P.~A.~Mazzali\altaffilmark{5,6}, S.~B.~Cenko\altaffilmark{7,8}, M.~M.~Kasliwal\altaffilmark{3},  Y.~Cao\altaffilmark{3}, A.~Horesh\altaffilmark{2}, N.~Palliyaguru\altaffilmark{1}, D.~A.~Perley\altaffilmark{9}, R.~R. Laher\altaffilmark{10}, F.~Taddia\altaffilmark{11}, G.~Leloudas\altaffilmark{2,9}, K.~Maguire\altaffilmark{12}, P.~E.~Nugent\altaffilmark{13,14}, J.~Sollerman\altaffilmark{11}, M.~Sullivan\altaffilmark{15}}
 \altaffiltext{1}{Department of Physics, Texas Tech University, Box 41051, Lubbock, TX 79409-1051, USA. E-mail: alessandra.corsi@ttu.edu}
\altaffiltext{2}{Benoziyo Center for Astrophysics, Weizmann Institute of Science, 76100 Rehovot, Israel.}
\altaffiltext{3}{Division of Physics, Mathematics, and Astronomy, California Institute of Technology, Pasadena, CA 91125, USA.}
\altaffiltext{4}{National Radio Astronomy Observatory, P.O. Box O, Socorro, NM 87801, USA}
\altaffiltext{5}{Astrophysics Research Institute, Liverpool John Moores University, Liverpool L3 5RF, UK.}
\altaffiltext{6}{Max-Planck Institut fur Astrophysik, Karl-Schwarzschildstr. 1, D-85748 Garching, Germany.}
\altaffiltext{7}{Astrophysics Science Division, NASA Goddard Space Flight Center, Mail Code 661, Greenbelt, MD 20771, USA.}
\altaffiltext{8}{Joint Space-Science Institute, University of Maryland, College Park, MD 20742, USA.}
\altaffiltext{9}{Dark Cosmology Centre, Niels Bohr Institute, University of Copenhagen, Juliane Maries Vej 30, 2100 Copenhagen, Denmark.}
\altaffiltext{10}{Spitzer Science Center, California Institute of Technology, M/S 314-6, Pasadena, CA 91125, USA}
\altaffiltext{11}{Department of Astronomy, The Oskar Klein Center, Stockholm University, AlbaNova, 10691 Stockholm, Sweden. }
\altaffiltext{12}{Astrophysics Research Centre, School of Mathematics and Physics, Queen's University Belfast, Belfast BT7 1NN, UK}
\altaffiltext{13}{Department of Astronomy, University of California, Berkeley, CA 94720-3411, USA.}
\altaffiltext{14}{Lawrence Berkeley National Laboratory, 1 Cyclotron Road, MS 50B-4206, Berkeley, CA 94720, USA}
\altaffiltext{15}{Department of Physics and Astronomy, University of Southampton, Southampton, SO17 1SX, UK.}

\begin{abstract}
Long duration $\gamma$-ray bursts are a rare subclass of 
stripped-envelope core-collapse supernovae that launch collimated relativistic outflows (jets).  All $\gamma$-ray-burst-associated 
supernovae are spectroscopically of Type Ic with broad lines, but the fraction of broad-lined Type Ic supernovae harboring low-luminosity $\gamma$-ray-burst remains largely unconstrained. Some supernovae should be accompanied by off-axis $\gamma$-ray burst jets that remain invisible initially, but then emerge as strong radio  sources (as the jets decelerate). However, this critical prediction of the jet model for $\gamma$-ray bursts has yet to be verified observationally. 
Here, we present K. G. Jansky Very Large Array observations 
of 15 broad-lined supernovae of Type Ic discovered by the Palomar Transient Factory in an untargeted manner. 
 Most of the supernovae in our sample exclude radio emission observationally similar to that of the radio-loud, relativistic SN\,1998bw. We constrain the fraction of 1998bw-like broad-lined Type Ic supernovae to be $\lesssim 41\%$ (99.865\% confidence). Most of the events in our sample also exclude off-axis jets similar to GRB\,031203 and GRB\,030329, but we cannot rule out off-axis $\gamma$-ray-bursts expanding in a low-density wind environment.  Three supernovae in our sample are detected in the radio. PTF11qcj and PTF14dby show late-time radio emission with average ejecta speeds of $\approx 0.3-0.4$\,c, on the dividing line between 
 relativistic and ``ordinary'' supernovae. The speed of PTF11cmh radio ejecta is poorly constrained.  We estimate that $\lesssim 85\%$ (99.865\% confidence) of the broad-lined Type Ic supernovae in our sample may harbor off-axis $\gamma$-ray-bursts expanding in media with densities in the range probed by this study.
 \end{abstract}


\section{Introduction}
\label{Introduction}
Long-duration ($T_{\gamma}\gtrsim 2$\,s) $\gamma$-ray bursts (GRBs) are extremely energetic explosions (typically, $\approx 10^{52}$\,erg released in $\approx 10$\,s, also referred to as collapsars) marking the deaths of massive stars \citep{Galama1998,Bloom2006}. According to the popular fireball model \citep{Piran2004,Meszaros2006}, the explosion launches relativistic jets in which magnetic fields are amplified and particles accelerated \citep{Rhoads1999}. Observers located within the initial jets' opening angle ($\theta_{j}\gtrsim \theta_{obs}$; ``on-axis'' observers) see an intense flash of $\gamma$-rays. Subsequent emission from the decelerating jets produces a (slowly) decaying broad-band afterglow. If the fireball model is correct, then off-axis GRBs should exist and be $\approx 2/\theta^2_j$ times more common than the ones we see in $\gamma$-rays \citep{Granot2002}. While $\gamma$-ray emission from off-axis GRBs cannot be observed, their longer-wavelength afterglow emission is expected to become observable at later times, once  the jet decelerates and starts spreading \citep{Nakar2002}. 

Off-axis GRBs have not been discovered so far, but in the light of the well-established connection between long-duration GRBs and core-collapse supernovae (SNe) of spectral type Ic with broad-lines \citep[BL-Ic;][]{Bloom2006}, a natural way to search for off-axis events is to observe this type of SNe and wait for the decelerating jet to emerge. While the SN optical emission traces the slower explosion debris ($v\approx 0.03-0.1$\,c), synchrotron emission from the fastest ejecta peaks in the radio band. There can be two major sources of radio emission associated with GRB-SNe, 
\begin{description}
\item[(i)] the SN shock, whose radio emission is brighter and earlier-peaking the faster the SN ejecta, with expected luminosity  and peak time of $\approx 10^{29}$\,erg\,cm$^{-2}$\,s$^{-1}$\,Hz$^{-1}$ and $\approx 10-30$\,d since explosion, respectively, for relativistic events like SN\,1998bw \citep[][]{Galama1998,Kulkarni1998,Berger2003}; 
\item[(ii)] the GRB jet which, if off-axis, would be observed only when the SN ejecta decelerate to mildly or sub-relativistic speeds, thus producing a delayed and nearly-isotropized radio emission. \end{description}

Radio is indeed the best wavelength range for identifying relativistic events such as SN\,1998bw, and/or off-axis GRBs \citep{Granot2003,Paczynski2001}. In the past, hundreds of SNe Ib/c have been targeted with the Karl G. Jansky Very Large Array \citep[VLA;][]{Berger2003,Soderberg2006,Bietenholz2013} and the fraction of SNe Ib/c associated with GRBs has been constrained to $\lesssim 1-3$\%. However, only a very small fraction of the Ic SNe targeted by these past studies were broad-lined, the only type of SNe observationally linked to GRBs. Moreover, many of these SNe Ib/c  were located in large, massive, and metal-rich hosts, while GRBs are rarely seen in such galaxies \citep[e.g.,][]{Modjaz2008,Levesque2010,Hjorth2012,Graham2013,Perley2013,Xu2013,Kelly2014,Kruhler2015,Perley2015}. Thus, the fraction of purely BL-Ic SNe harboring relativistic jets remained, observationally, largely unconstrained. 

Here, we present a sample of 15 SNe discovered by the Palomar Transient Factory and/or intermediate Palomar Transient Factory \citep[PTF/iPTF, hereafter we use PTF for simplicity;][]{Law2009,Rau2009}, optimized to search for off-axis GRBs, namely, a sample of BL-Ics selected blindly in random galaxies (mostly dwarfs). While all (long-soft) GRBs may be accompanied by BL-Ic SNe, not all of these SNe make GRBs \citep[e.g.,][]{Ofek2007,Soderberg2010,Milisavljevic2015}. Why should some stars follow a different  path to death, ending their lives as collapsars rather than as ``ordinary'' SNe, is still a mystery. This study aims at providing additional clues to help gain deeper insight into the nature of collapsar events. 

After collecting photometric data (Section \ref{photometry}) and classifying the SNe in our sample as belonging to the family of BL-Ic SNe (Section \ref{Spectralclassification}), we performed X-ray follow-up observations for some of the events (Section \ref{sec:XRT}) in search for X-ray signatures (not accompanied by $\gamma$-rays) from GRBs observed slightly off-axis (and/or ``dirty'' fireballs; Section \ref{sec:XRTconstraints}). We performed cm-wavelength follow-up observations of all the SNe in our sample with the VLA (Section \ref{sec:radio}), in search for SN\,1998bw-like radio emission (point (i) above) and/or later-time signatures of off-axis jets (point (ii) above). Our sample greatly enlarges the sample of radio-monitored BL-Ic SNe published over the last $\approx 10$ years \citep{Berger2003,Soderberg2006,Ghirlanda2013,Bietenholz2013}, and our observational strategy allows us to probe a portion of the radio luminosity-time since explosion phase space that was left largely unexplored by previous studies (Section \ref{1998bw-likeconstraints}). We constrain the portion of the explosion energy-wind/ISM density parameter space that is excluded under the hypothesis that GRB jets significantly off-axis ($\theta_j\approx 90$\,deg)  are associated with the SNe in our sample (Section \ref{offaxisconstraints}), and set an upper-limit on the fraction of BL-Ic SNe in our sample that show radio emission possibly compatible with off-axis GRBs expanding in media with densities in the range probed by this study (Section \ref{radiomodel}). Finally, we give our conclusions (Section \ref{conclusion}).

Hereafter,  we adopt cosmological parameter values of $H_0 = 69.6$\,km\,s$^{-1}$\,Mpc$^{-1}$, $\Omega_M = 0.286$, $\Omega_{\Lambda}= 0.714$ \citep{Wright2006,Bennett2014}. 

\begin{footnotesize}
\begin{center}
\begin{deluxetable*}{llllllll}
\tablecaption{BL-Ic SNe with VLA observations in our sample. Our sample includes a total of 15 SNe, 12 of which are presented here for the first time. PTF10bzf, PTF10qts, and PTF11qcj were previously discussed in \citet{Corsi2011}, \citet{Walker2014}, and \citet{Corsi2014} respectively.   \label{opticalTab}}
\tablehead{\colhead{PTF}  & \colhead{RA~Dec} & \colhead{$T_{\rm P48}$\tablenotemark{a}}& \colhead{$z$} & \colhead{$d_L$} & \colhead{$M_{R/g}$\tablenotemark{b}} &  \colhead{$v\,(\rm{Si})$} & \colhead{Ref.}\\
  \colhead{name}  &  \colhead{(J2000)}  &      \colhead{-}        &   \colhead{-} &  \colhead{-}       &  \colhead{[AB]}          &         \colhead{-}              &     \colhead{-}    \\
       \colhead{-}   &  \colhead{(hh:mm:ss~deg:mm:ss)}  & \colhead{(MJD)} &  &\colhead{(Mpc)} & \colhead{(mag)}   & \colhead{(km\,s$^{-1}$)} &  }\\
\endhead
\startdata
10bzf & 11:44:02.99 +55:41:27.6 &55250.504 &0.0498 &   223   & -18.3 &  $2.6\times10^4$&\citet{Corsi2011}\\
10qts & 16:41:37.60 +28:58:21.1 & 55413.260 & 0.0907 & 418 &  -19.4&    $1.7\times10^4$& \citet{Walker2014} \\
10xem & 01:47:06.88 +13:56:28.8 & 55470.340 & 0.0567 & 255 & -18.6 &    $2.0\times10^4$& This paper \\
10aavz & 11:20:13.36 +03:44:45.2 & 55514.485 &0.062 & 280 &   -19.2  &   $1.5\times10^4$& This paper\\
11cmh & 13:10:21.74 +37:52:59.6 & 55673.336  &0.1055 &  491 &  -18.6 &   $1.6\times10^4$& This paper\\
11img & 17:34:36.30 +60:48:50.6 & 55755.408  &0.158 &  761&  -19.6 &  $1.5\times10^4$& This paper\\
11lbm & 23:48:03.20 +26:44:33.5 & 55793.259  & 0.039  & 173 &  -18.0 &   $1.5\times10^4$& This paper\\
11qcj & 13:13:41.51 +47:17:57.0&   55866.520\tablenotemark{c}  & 0.0287 & 124 & -18.0 &  $1.2\times10^4$         & \citet{Corsi2014}\\
12as & 10:01:34.05 +00:26:58.4 & 55925.298 &   0.033 &  146&   -17.5&   $2.2\times10^4$& This paper\\
13u & 15:58:51.21 +18:13:53.1 & 56324.481 &   0.10 &  463& -18.9 &   $1.0\times10^4$& This paper\\
13alq\tablenotemark{d} & 11:48:02.09 +54:34:38.2 & 56394.359 &   0.054 &  242& -18.9  &  $ 2.3\times10^4$& \citet{Drake2013}\\
13ebw & 08:17:15.88 +56:34:41.6 & 56621.389 & 0.069 &  313&  -18.2&   $2.3\times10^4$& This paper\\
14dby & 15:17:06.29 +25:21:11.4 & 56832.238 &  0.074 &  337& -17.9&   $1.4\times10^4$& This paper\\
14gaq &  21:32:54.08 +17:44:35.6  &  56924.213    & 0.0826  &  378&    -18.0 & $1.9\times10^4$& This paper\\
15dld\tablenotemark{e} &  00:58:13.28 -03:39:50.3 & 57318.322 & 0.047 & 210 & -17.9 &$1.0\times10^4$& This paper/LaSilla-QUEST
\enddata
\tablenotetext{a}{Discoveries times (T$_{\rm P48}$) are from P48 observations in $R$-band,  except for the case of PTF14gaq which was discovered and observed with P48 in $g$-band.} 
\tablenotetext{b}{ $M_{R/g}$ is the maximum absolute magnitude in the P48 $R$-band for all of the SNe but PTF14gaq, for which  $M_{R/g}$ is measured in the P48 $g$-band. These magnitudes are corrected for galactic extinction \citep{Schlafly2011}.}
\tablenotetext{c}{The SN was visible in a previous $g$-band image taken on 2011 October 23.}
\tablenotetext{d}{a.k.a. CSS130415:114802+543439/SN\,2013bn: PTF13alq was also discovered by the CRTS \citep{Drake2009} and classified as a 1998bw-like type Ic SN by the Copernico Telescope in Asiago \citep{Tomasella2013}. }
\tablenotetext{e}{PTF15dld/LSQ15bfp was also discovered by the La Silla-QUEST variability survey \citep{LaSilla} and classified by the Public ESO Spectroscopic Survey of Transient Objects \citep[PESSTO;][]{Smartt2015}.}
\end{deluxetable*}
\end{center}
\end{footnotesize}

\section{The BL-Ic supernova sample}
\label{sample}
\subsection{P48 discovery and photometry}
\label{photometry}
$R$-band (or $g$-band) discoveries (and follow-up) of the SNe in our sample (Table \ref{opticalTab}) were obtained using the 48-inch Samuel Oschin telescope at the Palomar Observatory (P48), which is routinely used by the PTF/iPTF. Processed images were downloaded from the Infrared Processing and Analysis Center (IPAC) PTF archive \citep{Laher2014}. Photometry was performed relative to the SDSS $r$-band (or $g$-band) magnitudes of stars in the field \citep{York2000}. We used our custom pipeline which performs image subtraction, and then point spread function (PSF) photometry on stacks of PTF images extracted from the IPAC archive \citep[][]{Kate2012,Ofek2012,Ofek2013a}. The flux residuals from individual subtracted images were binned, and converted to magnitudes. Errors were estimated from the standard deviation of the photometric measurements in each bin. 

\begin{figure*}
\begin{center}
\includegraphics[width=14cm,angle=-90]{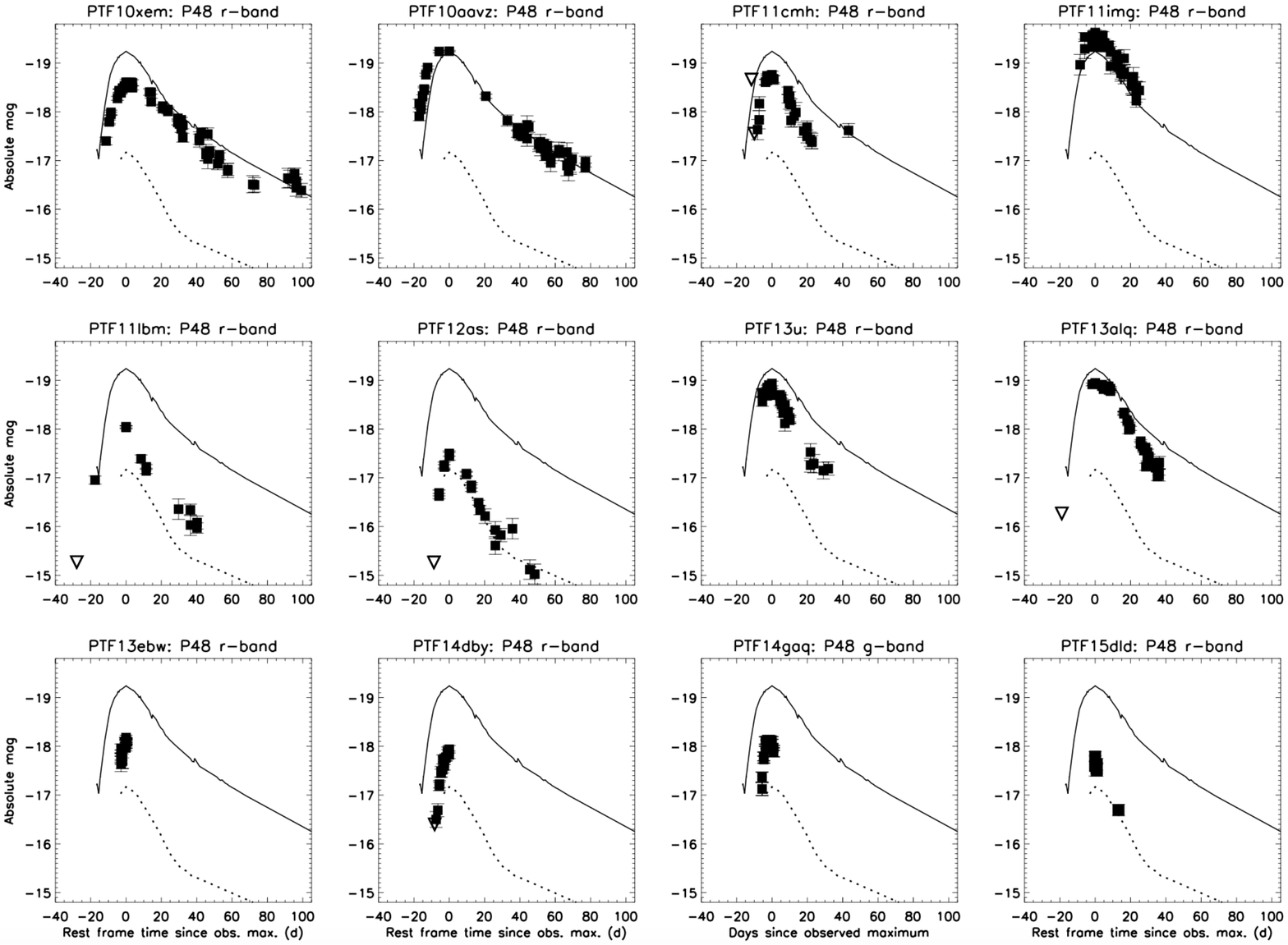}
\caption{The P48 $R$- or $g$-band light curves (corrected for Galactic extinction) of the BL-Ic SNe in our sample. PTF names are reported in the title of each panel. For comparison, we also show the $r$-band light curve of the GRB-associated BL-Ic SN\,1998bw \citep[solid line;][]{Clocchiatti2011}, and of the ``ordinary'' BL-Ic SN 2002ap \citep[dotted line;][]{Pandey2003}.  Epochs on the x-axis are measured since the time of maximum emission as observed by P48 (and corrected for redshift effects). Note that the time of maximum as observed by P48 is different from the SN light curve peak time for cases in which the peak emission was not observed by P48. \label{BLIc:light}}
\end{center}
\end{figure*}

The $R$-band (or $g$-band)  light curves of the SNe in our sample are shown in Fig. \ref{BLIc:light}. PTF10bzf, PTF10qts, and PTF11qcj photometry was discussed previously in \citet{Corsi2011}, \citet{Walker2014}, and \citet{Corsi2014} respectively, so we do not present their photometry here (we refer the reader to these papers). The P48 discovery time ($T_{P48}$), and the maximum $R$-band (or $g$-band) absolute magnitudes ($M_{R/g}$) as measured by our P48 monitoring and corrected for Galactic extinction \citep{Schlafly2011}, are reported in Table \ref{opticalTab}.  Note that our $M_{R/g}$ is different from the SN light curve peak for cases in which the peak emission was not observed by P48. We do not take into account k-corrections when measuring $M_{r/g}$, but refer the reader to \citet{Prentice2016} and Taddia et al. 2015 (in prep.) for a discussion of these corrections.

\begin{center}
\begin{figure*}
\includegraphics[width=14cm,angle=-90]{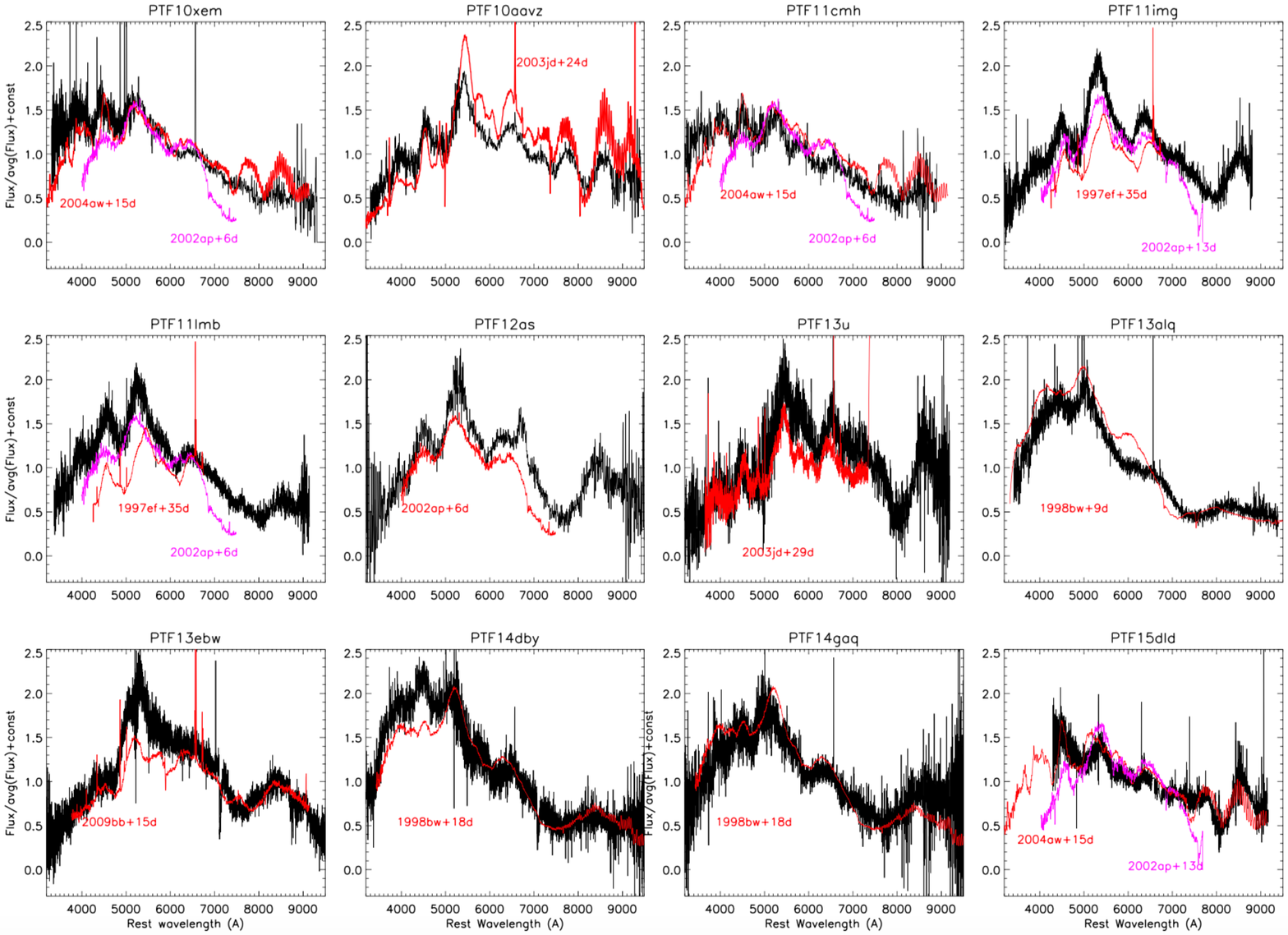}
\caption{Spectra of the SNe BL-Ic in our sample (black) compared to the spectra of the of the hypernova SN\,1997ef \citep[][epoch calculated since 1997 November 15]{Iwamoto1998,Branch1999}, of the GRB-associated BL-Ic SN\,1998bw \citep[epoch calculated since 1998 April 25;][]{Patat2001}, of the BL-Ic SN\,2002ap \citep[epoch calculated since 2002 January 28; ][]{GalYam2002,Mazzali2002}, of the BL-Ic/hyper-energetic and asymmetric SN\,2003jd \citep[epoch calculated since 2003 October 21; ][]{Valenti2008,Mazzali2005,Soderberg2006}, of the Ic/BL-Ic SN\,2004aw \citep[epoch calculated assuming an explosion date of $\approx 15$\,d before maximum; ][]{Taubenberger2006}, and of the relativistic BL-Ic SN\,2009bb \citep[epoch calculated since 2009 March 19; ][]{Soderberg2010,Pignata2011}. (See the electronic version of this paper for colors.) \label{BLIc:spectra}}
\end{figure*}
\end{center}
\subsection{Spectral classification}
\label{Spectralclassification}
After the discovery with P48, we triggered a spectroscopic follow-up campaign\footnote{All spectra reported in this work will be made public via WISeREP \citep{Yaron2012}.} of all the SNe in our sample. PTF10bzf, PTF10qts, and PTF11qcj spectral properties were previously discussed in \citet{Corsi2011}, \citet{Walker2014}, and \citet{Corsi2014} respectively, so we do not present their spectral analysis here, but we refer the reader to these papers. For the rest of the SNe in our sample, details of the observations are reported in what follows. In Table \ref{opticalTab}, we also report the estimated redshifts, and the velocities corresponding to the P-Cygni absorption minimum of the \ion{Si}{2} 6355\,\AA\ lines, that trace reasonably closely the position of the photosphere \citep[][see also Taddia et al. 2015, in prep.]{Mazzali2000}.
\subsubsection{PTF10xem}
On 2010 October 10 UT ($\approx 9$\,d since optical discovery), we observed PTF10xem with the dual-arm Kast spectrograph \citep{Miller1993} on the 3\,m Shane telescope at Lick Observatory. We used a 2\arcsec wide slit, a 600/4310 grism on the blue side, and a 300/7500 grating on the red side. Exposure time and air mass were 3600\,s and 1.09, respectively. The derived spectrum shows a good match with the Ic/BL-Ic SN\,2004aw \citep[e.g.,][]{Taubenberger2006} at an epoch of about 15\,d since explosion, and with the BL-Ic SN\,2002ap at  6\,d since explosion (Fig. \ref{BLIc:spectra}), so we classify PTF10xem as a BL-Ic SN.
\subsubsection{PTF10aavz}
On 2010 November 30 UT ($\approx 16$\,d since optical discovery) we observed PTF10aavz using ISIS on the William Herschel Telescope (WHT), with a 1.99\arcsec wide slit, the R300B grating set at a central wavelength of $\approx 4500$\,\AA\  on the blue side, and the R158R grating set at a central  wavelength of $\approx 7500$\,\AA\ on the red side. The exposure time was 1800\,s, and mean air mass was 1.17. This spectrum of PTF10aavz is most similar to the of the BL-Ic/hyper-energetic and asymmetric SN\,2003jd \citep[e.g.,][]{Valenti2008,Mazzali2005} at an epoch of about 24\,d since explosion (Fig. \ref{BLIc:spectra}).
\subsubsection{PTF11cmh}
We observed PTF11cmh using ISIS on the WHT on 2011 May 2 UT ($\approx 10$\,d since optical discovery), with a 1.02\arcsec wide slit, the R300B grating set at a central  wavelength of $\approx 4500$\,\AA\ on the blue side, and the R158R grating set at a central  wavelength of $\approx 7500$\,\AA\ on the red side. For both the blue and red side observations, the exposure time was 900\,s and the mean air mass was 1.01. The derived spectrum shows a good match with the Ic/BL-Ic SN\,2004aw \citep[e.g.,][]{Taubenberger2006} at an epoch of about 15\,d since explosion, and with the BL-Ic SN\,2002ap at  6\,d since explosion (Fig. \ref{BLIc:spectra}), so we classify PTF11cmh as a BL-Ic SN.
\subsubsection{PTF11img}
We observed PTF11img on 2011 August 2 UT ($\approx 20$\,d since optical discovery), using the Low Resolution Imaging Spectrometer \citep[LRIS;][]{LRIS} mounted on the Keck-I 10\,m telescope. The spectrum was taken using a 1\arcsec wide slit, with the 400/8500 grating set at a central  wavelength of $\approx 7800$\,\AA\ on the red side, and the 600/4000 grism on the blue side. For both sides, the exposure time and airmass were 600\,s and 1.46, respectively. The derived spectrum shows a good match with both the BL-Ic SN\,2002ap \citep[e.g.,][]{GalYam2002,Mazzali2002} at 13\,d since explosion, and the type Ic hypernova SN\,1997ef \citep[e.g.,][]{Iwamoto1998}  at 35\,d since explosion (Fig. \ref{BLIc:spectra}). We thus classify PTF11img as a BL-Ic SN.  
\subsubsection{PTF11lbm}
We observed PTF11lbm using ISIS on the WHT on 2011 August 31 UT ($\approx 11$\,d since optical discovery), with a 1.02\arcsec wide slit, the R300B grating set at a central  wavelength of $\approx 4500$\AA\ on the blue side; the R158R grating set at a central  wavelength of $\approx 7500$\,\AA\ on the red side. For both the blue and red side observations, the exposure time was 900\,s. The mean air mass was about 1.01. The derived spectrum shows a good match with both the BL-Ic SN\,2002ap \citep[e.g.,][]{GalYam2002,Mazzali2002} at 6\,d since explosion, and the type Ic hypernova SN\,1997ef \citep[e.g.,][]{Iwamoto1998} at 35\,d since explosion (Fig. \ref{BLIc:spectra}). We thus classify PTF11lbm as a BL-Ic SN.  
\subsubsection{PTF12as}
We observed PTF12as on 2012 January 2 UT ($\approx 2$\,d since optical discovery), using the Dual Imaging spectrograph (DIS) mounted on the 3.5\,m telescope at the Apache Point Observatory. The spectrum was taken using a 1.5\arcsec wide slit, with a B400/R300 grating setup. The exposure time and airmass were 1000\,s and 1.50, respectively. The derived spectrum shows a good match with the GRB-associated BL-Ic SN\,2002ap \citep[e.g.,][]{Mazzali2002,GalYam2002} at 6\,d since explosion (Fig. \ref{BLIc:spectra}).
\subsubsection{PTF13u}
On 2013 February 18 UT ($\approx 17$\,d since optical discovery) we observed PTF13u with the Double Beam Spectrograph \citep[DBSP;][]{Oke1982} on the Palomar 200-inch telescope (P200). We used the 316/7500 and 600/4000 gratings for red and blue camera respectively, with a D55 dichroic, resulting in a spectral coverage of $\approx (3500-9500)$\,\AA. Exposure times and air mass were of 590\,s and 1.06, respectively. This spectrum of PTF13u matches that of  the BL-Ic/hyper-energetic and asymmetric SN\,2003jd \citep[e.g.,][]{Valenti2008,Mazzali2005} at $\approx 29$\,d since explosion (Fig. \ref{BLIc:spectra}). 
\subsubsection{PTF13alq}
On 2013 April 13 UT ($\approx 1$\,d since optical discovery) we observed PTF13alq with the DBSP \citep[][]{Oke1982} on P200. We used the 316/7500 and 600/4000 gratings for red and blue camera respectively, with a D55 dichroic, resulting in a spectral coverage of $\approx (3500-9500)$\,\AA. Exposure times and air mass were of 300\,s and 1.1, respectively. The derived spectrum shows a good match with the GRB-associated BL-Ic SN\,1998bw \citep[e.g.,][]{Patat2001} at 9\,d since explosion (Fig. \ref{BLIc:spectra}).
\subsubsection{PTF13ebw}
We observed PTF13ebw on 2013 December 4 UT ($\approx 9$\,d since optical discovery), using LRIS mounted on the Keck-I 10\,m telescope. The spectrum was taken using a 1\arcsec wide slit, with the 400/8500 grating set at a central wavelength of $\approx 7800$\,\AA\ on the red side, and the 600/4000 grism on the blue side. The exposure time and airmass were 500\,s and 1.36, respectively. The derived spectrum shows a good match with the relativistic BL-Ic SN\,2009bb \citep{Soderberg2010} at 15\,d since explosion (Fig. \ref{BLIc:spectra}).
\subsubsection{PTF14dby}
On 2014 June 29 ($\approx 5$\,d since optical discovery), we observed PTF14dby with LRIS mounted on the Keck-I 10\,m telescope. The spectrum was taken using a 1\arcsec wide slit, with the 400/8500 grating set at a central wavelength of $\approx 7800$\,\AA\ on the red side, and the 400/3400 grism on the blue side. The exposure time and airmass were 300\,s and 1.19, respectively. This spectrum of PTF14dby reveals a good match with the GRB-associated BL-Ic SN\,1998bw \citep[e.g.,][]{Patat2001} at $\approx 18$\,d since explosion (Fig. \ref{BLIc:spectra}). 
\subsubsection{PTF14gaq}
We observed PTF14gaq on 2014 October 1 UT ($\approx 7$\,d since optical discovery) with the DBSP on P200. We used the 316/7500 and 600/4000 gratings for the red and blue camera respectively, with a D55 dichroic, resulting in a spectral coverage of $\approx (3500-9500)$\,\AA. Exposure times and air mass were of 600\,s and 1.05, respectively. This spectrum of PTF14gaq shows a good match with the GRB-associated SN\,1998bw \citep[e.g.,][]{Patat2001} at $\approx 18$\,d since explosion (Fig. \ref{BLIc:spectra}). 
\subsubsection{PTF15dld}
We observed PTF15dld on 2015 November 7 UT ($\approx 15$\,d since the P48 optical discovery) with the Deep Extragalactic Imaging Multi-Object Spectrograph (DEIMOS) mounted on the Keck-II 10\,m telescope. The spectrum was taken using the 600ZD grating and GG455 filter. The exposure time and airmass were 600\,s and 1.11, respectively. The derived spectrum shows a good match with both the BL-Ic SN\,2002ap \citep[e.g.,][]{GalYam2002,Mazzali2002} at 13\,d since explosion, and with the Ic/BL-Ic SN\,2004aw \citep[e.g.,][]{Taubenberger2006} at an epoch of about 15\,d since explosion (Fig. \ref{BLIc:spectra}).

\begin{footnotesize}
\begin{center}
\begin{deluxetable*}{llllllllll}
\tablecaption{$3\sigma$ X-ray upper-limits or detections for some of the SNe in our sample.  \label{X}}
\tablehead{\colhead{PTF} & \colhead{Date} & \colhead{$\Delta T_X$\tablenotemark{f}} & \colhead{Instrument} & \colhead{Band} & \colhead{Exp.} & \colhead{$N_H$\tablenotemark{g}}  & \colhead{Count Rate} & \colhead{Flux (unabs)\tablenotemark{h}} & \colhead{Ref.}\\
  \colhead{name} &\colhead{(MJD)} &      \colhead{(days)}&  \colhead{-}    & \colhead{(keV)} & \colhead{(ks)}     & \colhead{($10^{20}$\,cm$^{-2}$)} & \colhead{($10^{-4}$\,s$^{-1}$)} & \colhead{($10^{-14}$ erg\,cm$^{-2}$\,s$^{-1}$)}}\\
\endhead
\startdata
10bzf & 55259.290  & 9 &  \textit{Swift}-XRT & 0.3-10 & 5.0  &   0.88  &  $<3.7$  &   $<1.3$   & \cite{ATEL2471}\\
''    &   55263.303  &  13 &  \textit{Swift}-XRT & 0.3-10 & 5.0  &   "  &  $<7.7$  &   $<2.7$   & \cite{ATEL2471}\\
10qts & 55426.126 & 13&\textit{Swift}-XRT & 0.3-10 & 4.9 & 2.7 & $<4.2$ & $<1.6$ & This paper\\
11qcj & 55920.045 & 78 & \textit{Swift}-XRT & 0.3-10 &  31.4& 1.0 & $<4.1$& $<1.4$ & \citet{Corsi2014}\\
'' &    55939.05  &  97   & \textit{Chandra}-ACIS & 0.3-8.0 &9.8   &  ''     &    $8.8\pm3.3$\tablenotemark{i}   & $0.76\pm0.26$   & \cite{Corsi2014}            \\
12as  & 55932.340 & 7 & \textit{Swift}-XRT & 0.3-10 & 4.6 & 2.5 & $<33$ & $<13$ & This paper\\
''    & 55962.150 & 37 &\textit{Swift}-XRT & 0.3-10 & 4.9 & ''  & $<21$ & $<8.0$ & This paper\\
13ebw & 56631.719 & 10 &\textit{Swift}-XRT & 0.3-10 & 4.8 & 4.6 & $<24$ & $<9.8$ & This paper\\
''    & 56672.486 & 51 & \textit{Swift}-XRT & 0.3-10 & 2.4 & '' & $<48$ & $<20$ & This paper\\
14dby      &  56839.043 & 7 & \textit{Swift}-XRT & 0.3-10 & 9.5 &  4.2 & $<11$ & $<4.4$ & This paper\\
14gaq      &  56936.081& 12 & \textit{Swift}-XRT & 0.3-10 & 7.8 &  6.9& $<14$ & $<6.1$ & This paper\\
15dld       &  57332.973  &  15     & \textit{Swift}-XRT & 0.3-10 &   9.9   &    3.2     & $<9.8$ &   $<3.8$    & This paper
\enddata
\tablenotetext{f}{Epoch in days since P48 discovery (see Table 1), not corrected for redshift effects.}
\tablenotetext{g}{Hydrogen column densities are weighted averages from the Leiden/Argentine/Bonn (LAB) Survey of Galactic \ion{H}{1} \citep{Kalberla2005}.}
\tablenotetext{h}{The count-rate-to-flux conversion assumes a photon index of $\Gamma_X=2$.}
\tablenotetext{i}{\textit{Chandra} observations of PTF11qcj yielded detections which we attribute to the presence of strong CSM interaction rather than to a GRB X-ray afterglow. Some host galaxy contamination to the measured X-ray flux might also be present. See \citet{Corsi2014} for a complete discussion. }
\end{deluxetable*}
\end{center}
\end{footnotesize}

\subsection{\textit{Swift/XRT} follow-up and data reduction}
\label{sec:XRT}
None of the BL-Ic SNe in our sample was found to be spatially coincident with any of the well-localized GRB in the \textit{Swift} \citep{Gehrels2004} catalog. For some of the events, we triggered \textit{Swift/XRT} \citep{Burrows2005} follow-up observations via our approved Target of Opportunity Programs\footnote{Program IDs 1013248 and 1114155 (PI: Corsi).} in order to further exclude the presence of a GRB X-ray afterglow with no associated $\gamma$-rays \citep[as would be the case for a GRB jet observed slightly off-axis, or for a so-called ``dirty'' fireball; see e.g.][]{Rhoads2003}. 

We downloaded the \textit{Swift}-XRT data from the archive\footnote{See \url{http://heasarc.gsfc.nasa.gov/cgi-bin/W3Browse/swift.pl}.}. None of the SNe in our sample yielded a detection with  \textit{Swift/XRT}, so we calculated 3$\sigma$ upper-limits on the 0.3-10.0\,keV count rate using standard analysis procedures.  The upper-limits are reported in Table \ref{X}, where we have converted the 0.3-10\,keV XRT count rates into fluxes assuming a photon index of $\Gamma_X=2$ and correcting for Galactic absorption.

X-ray observations of PTF11qcj obtained with \textit{Swift}/XRT and \textit{Chandra}/ACIS \citep{Garmire2003} were previously presented in \citet{Corsi2014}. We include some of these observations (the most significant \textit{Chandra}/ACIS detection and the deepest \textit{Swift}/XRT upper-limit) in Table \ref{X} for completeness.

\begin{figure}
\begin{center}
\includegraphics[width=6.8cm,angle=-90]{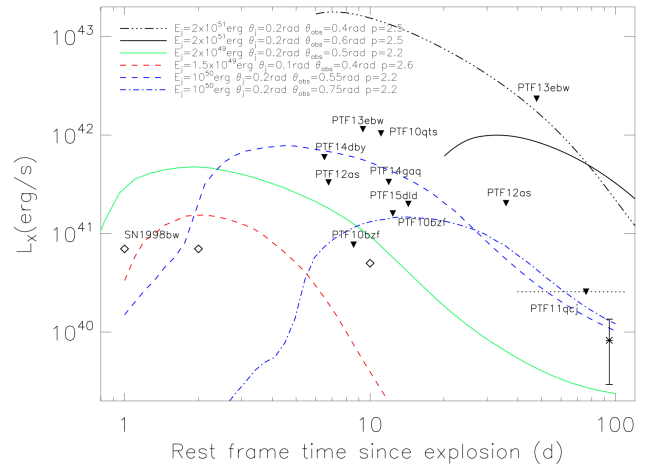}
\caption{\textit{Swift}/XRT upper-limits on some of the BL-Ic SNe in our sample (downward pointing triangles) compared with the X-ray emission from GRB\,980425 (diamonds) and with the X-ray emission expected from off-axis GRB models by \citet{van2011,van2012}. GRB jets opening angles are set to  $\theta_j=(0.1-0.2)$\,rad; the medium is a constant density ISM ($n_{\rm ISM}=1-10$\,cm$^{-3}$); observer's viewing angles are in the range $\theta_{obs}\approx (2-4)\theta_j$. The fraction of energy density of the ejecta going into electrons ($\epsilon_e$) and magnetic fields ($\epsilon_B$) are both set to 0.1 in all of the above models.  For PTF11qcj we plot the flux obtained from the most significant \textit{Chandra}/ACIS detection after subtracting the possible host galaxy contribution (asterisk), and the deepest \textit{Swift}/XRT upper limit (dotted line). We attribute the X-ray emission from PTF11qcj to the presence of strong CSM interaction rather than to a GRB X-ray afterglow. See \citet{Corsi2014} for a complete discussion. (See the electronic version of this paper for colors.) \label{X-rays}}
\end{center}
\end{figure}

\subsection{VLA follow-up observations and data reduction}
\label{sec:radio}
We observed all of the SNe in our sample, along with the necessary calibrators, with the VLA\footnote{The National Radio Astronomy Observatory is a facility of the National Science Foundation operated under cooperative agreement by Associated Universities, Inc.; \url{http://www.nrao.edu/index.php/about/facilities/vlaevla}} \citep[][]{Perley2009} under our Target of Opportunity programs\footnote{VLA/11A-227, VLA/11B-034, VLA/12B-247, VLA/14A-434, and VLA/15A-314 - PI: A. Corsi; and VLA/10B-221 - PI: Kasliwal).}. VLA data were reduced and imaged using the Common Astronomy Software Applications (CASA) package. 

The VLA flux measurements and/or upper-limits are reported in Table \ref{radioTab}. Measurement errors are calculated by adding in quadrature the rms map error, and a basic fractional error ($\approx 5\%$) which accounts for inaccuracies of the flux density calibration \citep{Weiler1986,Eran2011}. 

\section{X-ray constraints on associated GRB X-ray afterglows}
\label{sec:XRTconstraints}
As mentioned in the previous Section, none of the SNe in our sample are spatially coincident with any well-localized GRB. For a limited number of these SNe, our observations with the \textit{Swift}/XRT allow us to  make some comparisons with the X-ray light curve that would be expected from an accompanying GRB\,980425-like event, or from a high-luminosity GRB observed slightly off-axis.  For the last, we use the numerical model by \citet{van2011,van2012}, which considers a relativistic GRB fireball expanding in a uniform density medium. 

As evident from Fig. \ref{X-rays}, while \textit{Swift}/XRT upper-limits can exclude X-ray afterglows associated with high-luminosity (high-energy) GRBs observed slightly off-axis (up to $\theta_{obs}\lesssim (2-3)\theta_j$), X-ray emission as faint as the afterglow of the low-luminosity GRB\,980425 cannot be excluded. As we discuss in Section \ref{1998bw-likeconstraints}, radio data collected with the VLA enable us to exclude 1998bw-like emission for most of the SNe in our sample. 

For PTF11qcj, \textit{Chandra} observations yielded a detection but we attribute this X-ray emission to the presence of strong CSM interaction rather than to a GRB X-ray afterglow. See \citet{Corsi2014} for a complete discussion. 

\section{Constraining the fraction of 1998bw-like events using radio emission}
\label{1998bw-likeconstraints}
Here, we aim at observationally constraining the fraction of BL-Ic SNe with radio luminosities comparable to that of the GRB-associated SN\,1998bw \citep{Kulkarni1998} to ultimately constrain the fraction of BL-Ic SNe harboring low-luminosity GRBs. Indeed, most of the GRBs with an associated and spectroscopically confirmed SN are low-luminosity bursts ($E_{\gamma,iso}\lesssim 10^{50}$\,ergs), although notable exceptions are GRB\,030329 \citep{Stanek2003} and GRB\,130427A \citep[e.g.,][]{Melandri2014,Perley2014}. The fraction of BL-Ic SNe harboring low-luminosity GRBs was left largely unconstrained by previous efforts  \citep[][]{Berger2003,Soderberg2006,Bietenholz2013} due to the very small number of BL- Ic events with radio follow-up available to the community.  

Theoretical studies have indirectly constrained the fraction of BL-Ic SNe harboring low-luminosity GRBs by constraining the local rate of low-luminosity GRBs (via luminosity function fitting) and then comparing this estimated local rate with the rate of BL-Ic SNe collected via optical surveys.  Following this approach, \citet{Guetta2007} derived that $\gtrsim 10\%$ of BL-Ic SNe are accompanied by low-luminosity GRBs. This is consistent with the earlier results by \citet{Podsiadlowski2004}, who found that the rates of GRBs and BL-Ic SNe are comparable to within the uncertainties, and their ratio likely $\gtrsim 30\%$ \citep[see Table 1 in][]{Podsiadlowski2004}. More recently, following a statistical approach inspired by the Drake equation, \citet{Graham2015} estimated that there are $4000\pm2000$ BL-Ic SNe in low-metallicity environments for every (long) GRB aligned in our direction.  This number is a composite of the fraction of such SNe which produce GRBs and the fraction that are beamed in our direction, and would imply that $\lesssim 5\%$ of the BL-Ic SNe are associated with a GRB.

\begin{figure}
\begin{center}
\vspace{-0.5cm}
\includegraphics[width=9cm]{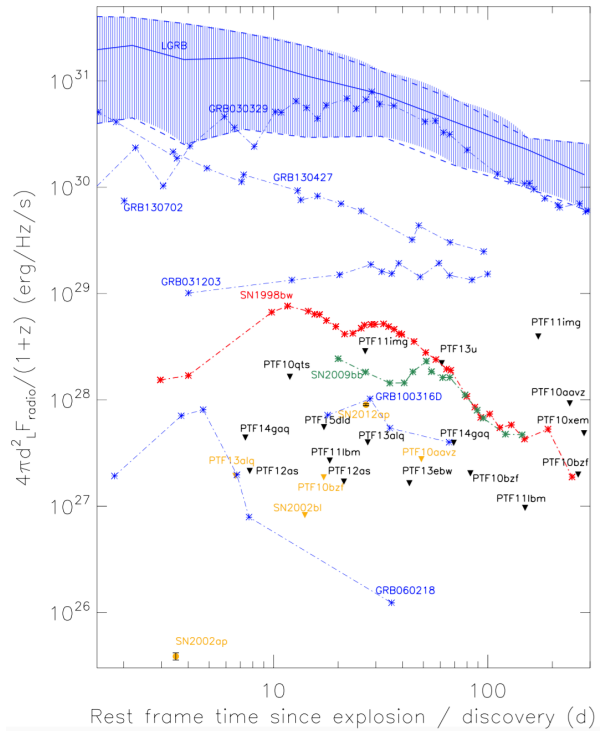}
\vspace{-1.2cm}
\caption{Radio (observed central frequencies of $\approx 4.8-6.3$\,GHz; see Table \ref{radioTab}) upper-limits for the BL-Ic SNe in our sample with VLA follow-up observations at $t\lesssim 300$\,d since discovery (as measured in the SN rest frame), compared with: the mean radio (8.5\,GHz) light curve of cosmological GRBs (blue solid line) as derived by \citet{Chandra2012}, together with the 75\% confidence interval (blue shaded region); radio ($\approx 5$\,GHz) afterglow light curves of long GRBs with spectroscopically-associated SNe \citep[blue asterisks;][]{Berger030329,Soderberg2004,Frail2005,Soderberg060218,Margutti2013,Perley2014,Singer2015}; the light curve of the GRB-SN\,1998bw \citep[red asterisks;][]{Kulkarni1998}; the light curve of the relativistic BL-Ic SN\,2009bb \citep[green asterisks;][]{Soderberg2010} and SN\,2012ap \citep[yellow dot;][]{Cha2015,Milisavljevic2015}. Black triangles are our upper-limits. Yellow data points or upper-limits are for BL-Ic SN radio ($\approx 5-8.5$\,GHz) follow-up observations that were not conducted via our programs \citep[][]{Berger2003,ATEL2483,Soderberg2010,ATEL3101,ATEL4997}. Radio detections for the PTF sample are plotted in Fig. \ref{radioint}. Late-Time radio observations of BL-Ic SNe performed at $t\gtrsim 300$\,d since explosion via other studies \citep{Soderberg2006,Bietenholz2013} are not reported here. \label{earlyradio}}
\end{center}
\end{figure}

\begin{figure}
\begin{center}
\includegraphics[width=6.5cm,angle=-90]{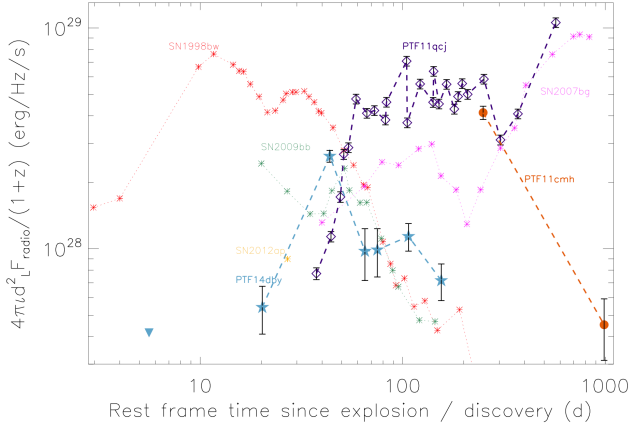}
\caption{BL-Ic SNe in our sample with radio detections: PTF11cmh, orange dots; PTF11qcj, purple diamonds \citep{Corsi2014}; PTF14dby, blue stars. We compare these non-relativistic / CSM-interacting SNe with the light curves of the GRB-SN\,1998bw \citep[red asterisks;][]{Kulkarni1998}, of the relativistic BL-Ic SN\,2009bb \citep[green asterisks;][]{Soderberg2010}, of the CSM-interacting BL-Ic SN\,2007bg \citep[magenta asterisks;][]{Salas2013}, and of the relativistic SN\,2012ap \citep[yellow asterisk;][]{Cha2015}. As evident from this comparison, the non-relativistic and CSM-interacting BL-Ic peak at later timescale than the relativistic ones. \label{radioint}}
\end{center}
\end{figure}

With our PTF discoveries, we now have a sample of 15 BL-Ic SNe discovered \textit{independently of a GRB trigger}, with at least one radio follow-up observation on timescales $\lesssim 300$\,d since explosion  (as measured in the SN rest frame; Figs. \ref{earlyradio} and \ref{radioint}). Of these 15 SNe, 12 have been uniquely observed via our VLA programs. Our observations have greatly enlarged the sample of 8 BL-Ic SNe with radio follow-up  at $\lesssim 300$\,d collected via independent efforts during the last decade \citep[Fig. \ref{earlyradio}, yellow; Fig. \ref{radioint}, green, yellow, and magenta asterisks;][]{Berger2003,ATEL2483,Soderberg2010,ATEL3101,Drake2013,ATEL4997,Salas2013,Cha2015,Milisavljevic2015}.  We are thus in a position to start constraining the theoretical expectations for the low-luminosity GRB-to-BL-Ic SN ratio using a direct observational signature: the presence (or absence) of 1998bw-like radio emission.

In Figs. \ref{earlyradio} and \ref{radioint}, the 5\,GHz radio light curve of SN\,1998bw is compared with the upper-limits and detections obtained for the SNe in our sample (4.8-6.3 GHz; See Table \ref{radioTab}). As evident from Fig. \ref{radioint}, we have 3 SNe (PTF11cmh, PTF11qcj, and PTF14dby) that show bright radio emission, much brighter than the ordinary BL-Ic SN\,2002ap and almost at the level of SN\,1998bw,  but their radio peak occurs $\gtrsim 5\times$ later than for SN\,1998bw. Thus, as we explain in the following Section, we consider these SNe as observationally different from SN\,1998bw, likely related to events on the dividing line between ordinary SNe and GRBs, although an interpretation as off-axis GRB jets might also be possible. For the remaining 12 SNe in our sample, we detect no radio emission and set upper-limits (Fig. \ref{earlyradio}, black downward-pointing triangles). For 10 of these 12 SNe (all but PTF10xem and PTF13u), we have at least one upper-limit which constrains the radio emission to be dimmer than the emission of SN\,1998bw \textit{at a similar epoch}, thus excluding a radio light curve observationally similar to that of the prototype relativistic BL-Ic SN\,1998bw.  \citep[We note that 8 out of the 12 SNe also exclude radio emission similar to SN\,2009bb, a relativistic  BL-Ic SN with no associated GRB;][]{Soderberg2010}.

Based on the above results, we conclude that of the 10+3 PTF SNe whose radio observations can set constraints on SN\,1998bw-like emission, none of them was in fact like SN\,1998bw in the radio i.e., they were all \textit{observationally} different. Adding to this sample  the BL-Ic SN\,2002ap \citep{GalYam2002,Mazzali2002} and SN\,2002bl \citep{Armstrong2004,Berger2003}, and the CSM-interacting BL-Ic SN\,2007bg \citep{Salas2013}, we have a total of 16 BL-Ic SNe for which radio emission observationally similar to SN\,1998bw is excluded. Because the 99.865\% confidence  ($3\sigma$ Gaussian equivalent for a single-sided distribution) Poisson upper-limit on zero SNe compatible with SN\,1998bw is $\approx 6.61$, we conclude that the rate of BL-Ic SNe observationally similar to SN\,1998bw $\lesssim 6.61/16\approx 41\%$.  

We note that for these 16 SNe we also exclude on-axis radio emission typical of long GRB afterglows at cosmological distances \citep[blue line and shaded area in Fig. \ref{earlyradio}; see ][]{Chandra2012}, and radio emission observationally similar to that of the low-luminosity GRB\,031203. However, none of our upper-limits exclude radio emission similar to GRB\,060218. This is not surprising since the afterglow of this low-luminosity GRB faded on timescales much faster than the ones our VLA monitoring campaign was designed to target (peak timescales of $\approx 20-30$\,d). Finally, only some of our upper-limits exclude radio afterglow emission similar to that of the low-luminosity GRB\,100316D (although a more quantitative comparison with this burst is hampered by its poorly sampled radio light curve). 

\section{Constraining the fraction of (largely) off-axis GRBs from radio non-detections}
 \label{offaxisconstraints}
 Low-luminosity GRBs (such as GRB\,980425 associated with SN\,1998bw) are believed to be intrinsically less energetic events (when compared to high-luminosity ones) with jet opening angles $\gtrsim 30$\,deg \citep[e.g.,][]{Liang2007}. However, the possibility that low-luminosity GRBs are higher-energy events observed off-axis has also been discussed \citep[e.g.,][]{Waxman2004,Ramirez2005}. Indeed, most (high-luminosity) GRBs are believed to have opening angles of the order of $\sim 10$\,deg \citep[e.g.,][]{Frail2001,Liang2008,Racusin2009,Zhang2015}.  Here, we aim at answering the question of whether the BL-Ic SNe in our sample that do not show evidence for radio emission observationally similar to that of SN\,1998bw (Section \ref{1998bw-likeconstraints}) could still be accompanied by an off-axis ($\theta_{obs}\approx 90$\,deg) GRB afterglow that would become visible in the radio band long past the explosion \citep[at timescales of the order of $\sim 1$\,yr;][]{Levinson2002,Waxman2004,Gal-Yam2006}, when the relativistic fireball enters the sub-relativistic phase and starts spreading, rapidly intersecting the viewer's line of sight while approaching spherical symmetry.  Our upper-limits add to the late-time  ones that have been collected in the past \citep[$t\gtrsim 500$\,d since explosion, not plotted in Figs. \ref{earlyradio}-\ref{radioint}; see][]{Soderberg2006,Bietenholz2013}.
  
   We model the late-time radio emission from an off-axis GRB following the works by \citet{Livio2000} and \citet{Waxman2004}, modified to account for the results of more recent numerical simulations \citep{Zhang2009,vanEerten2012}.  The last have shown that at the end of the Blandford-McKee  (BM) phase, the fireball becomes non relativistic but, differently from what previously thought, the transition to the spherical Sedov-Neumann-Taylor (SNT) blast wave takes a rather long time. Thus, accurate modeling of the fireball evolution over timescales in between the BM and SNT phases (which are relevant for this study) requires numerical simulations. However, \citet{Zhang2009} have shown that for fireballs expanding in a medium of constant density $n_{0,ISM}$ (in units of cm$^{-3}$) and at timescales
   \begin{equation}
  t\gtrsim (1+z)\times t_{\rm SNT}/2  , \label{timeconstr}
  \end{equation}
    where
  \begin{equation}
  t_{\rm SNT}\approx 92\,{\rm d} \left(\frac{E_{51}}{n_{0,ISM}}\right)^{1/3},\label{tSNT}
   \end{equation}
 an acceptable analytical approximation to the afterglow flux is given by:
 \begin{eqnarray}
\nonumber   F_{\nu}(t)\approx 0.16 \,d^{-2}_{L,28}(1+z)^{\frac{1}{2}}\left(\frac{\epsilon_e}{0.1}\right)\left(\frac{\epsilon_B}{0.1}\right)^{3/4}n_{0,ISM}^{\frac{9}{20}}E^{\frac{13}{10}}_{51}~~~\\ \times \left(\frac{\nu}{\rm 1\,GHz}\right)^{-1/2}\left(\frac{t}{92\,{\rm d}(1+z)}\right)^{-9/10}\,\rm{mJy}.~~~\label{eqZhang}
 \end{eqnarray}  
Here $E_{51}$ is the beaming-corrected ejecta energy in units of $10^{51}$ erg; $\epsilon_e$ and $\epsilon_B$ are the fraction of ejecta energy density going into electrons and magnetic fields, respectively; $d_{L,28}$ is the luminosity distance of the source in units of $10^{28}$\,cm; $z$ is the source redshift; and the power-law index of the electron energy distribution has been set to $p\approx 2$. 

Equation  (\ref{eqZhang}) corresponds to Eq. (15) in \citet{Waxman2004} where, however, the dependence on $t_{\rm SNT}$ has been eliminated by using our Eq. (\ref{tSNT}) \citep[or, equivalently, Eq. (11) in ][]{Waxman2004}. Using Eq.  (\ref{eqZhang}) to constrain the fireball parameters by comparison with observations taken at timescales $t$ that satisfy Eq. (\ref{timeconstr}),  yields constraints on the (beaming-corrected) energy that are accurate to within a factor of $\approx 2$ \citep[see Fig. 10 in][]{Zhang2009}.  

\begin{figure*}
\begin{center}
\hbox{
\includegraphics[width=5.cm,angle=-90]{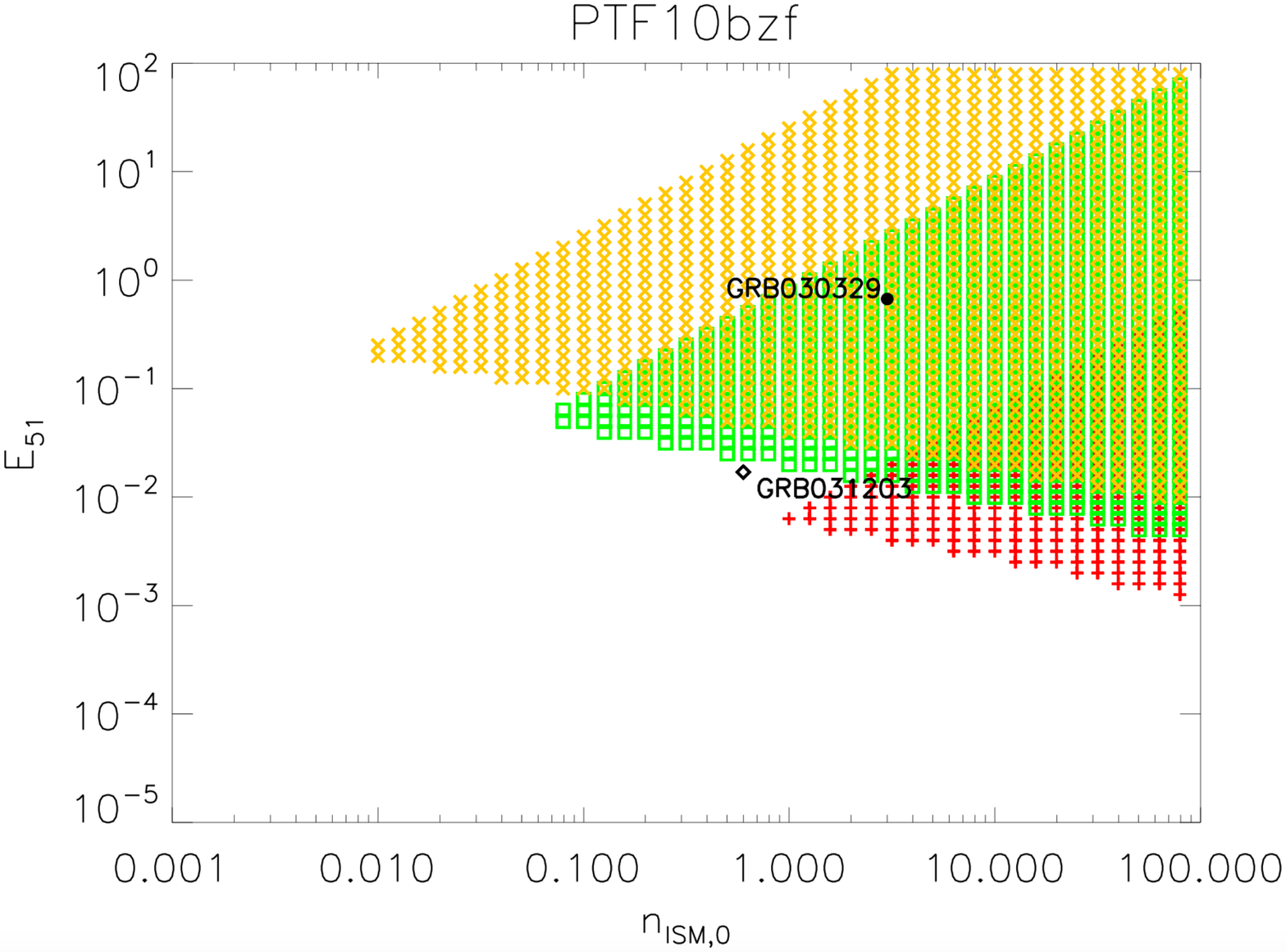}
\hspace{-0.2cm}
\includegraphics[width=5.cm,angle=-90]{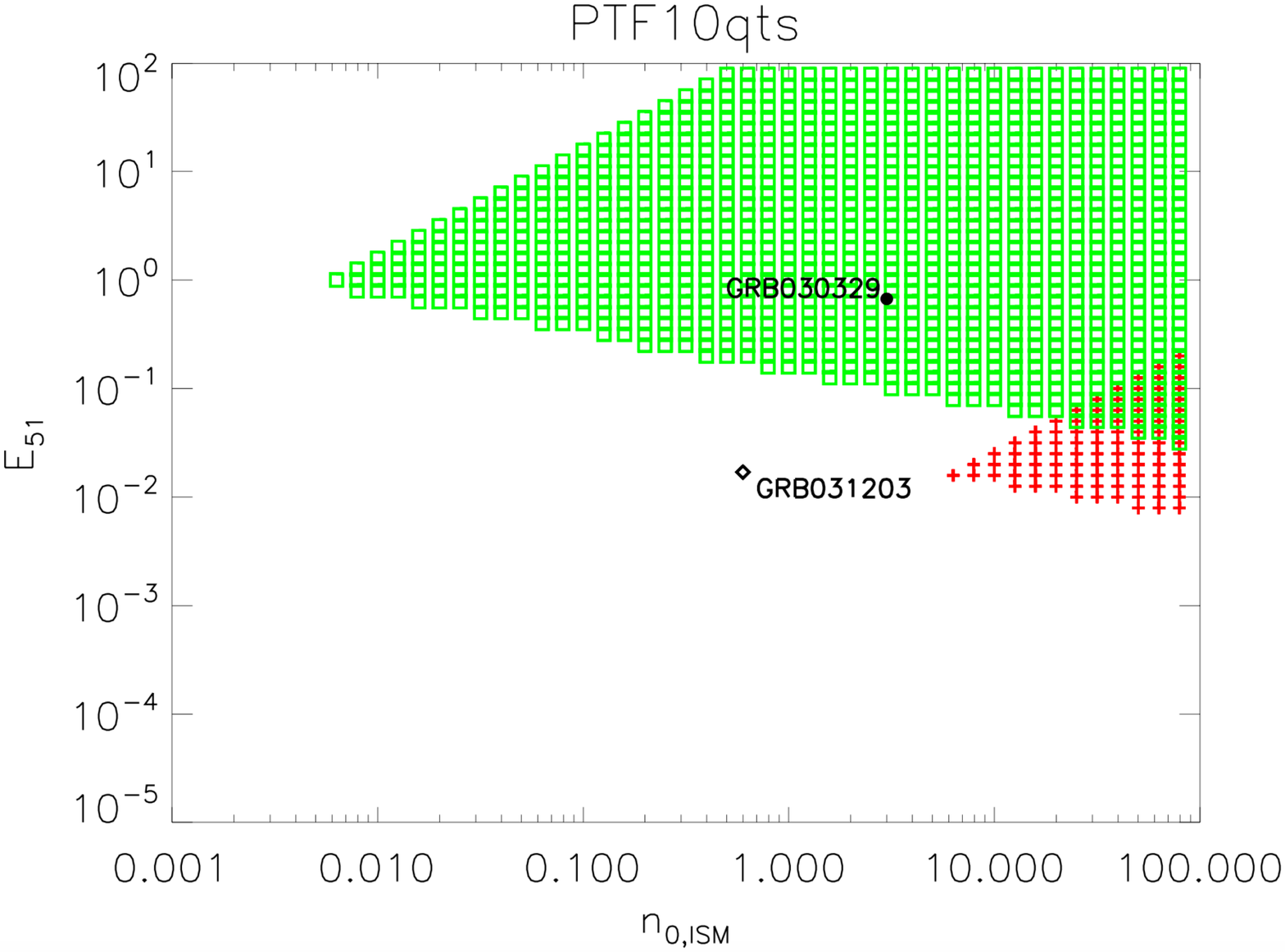}
\hspace{-0.2cm}
\includegraphics[width=5.cm,angle=-90]{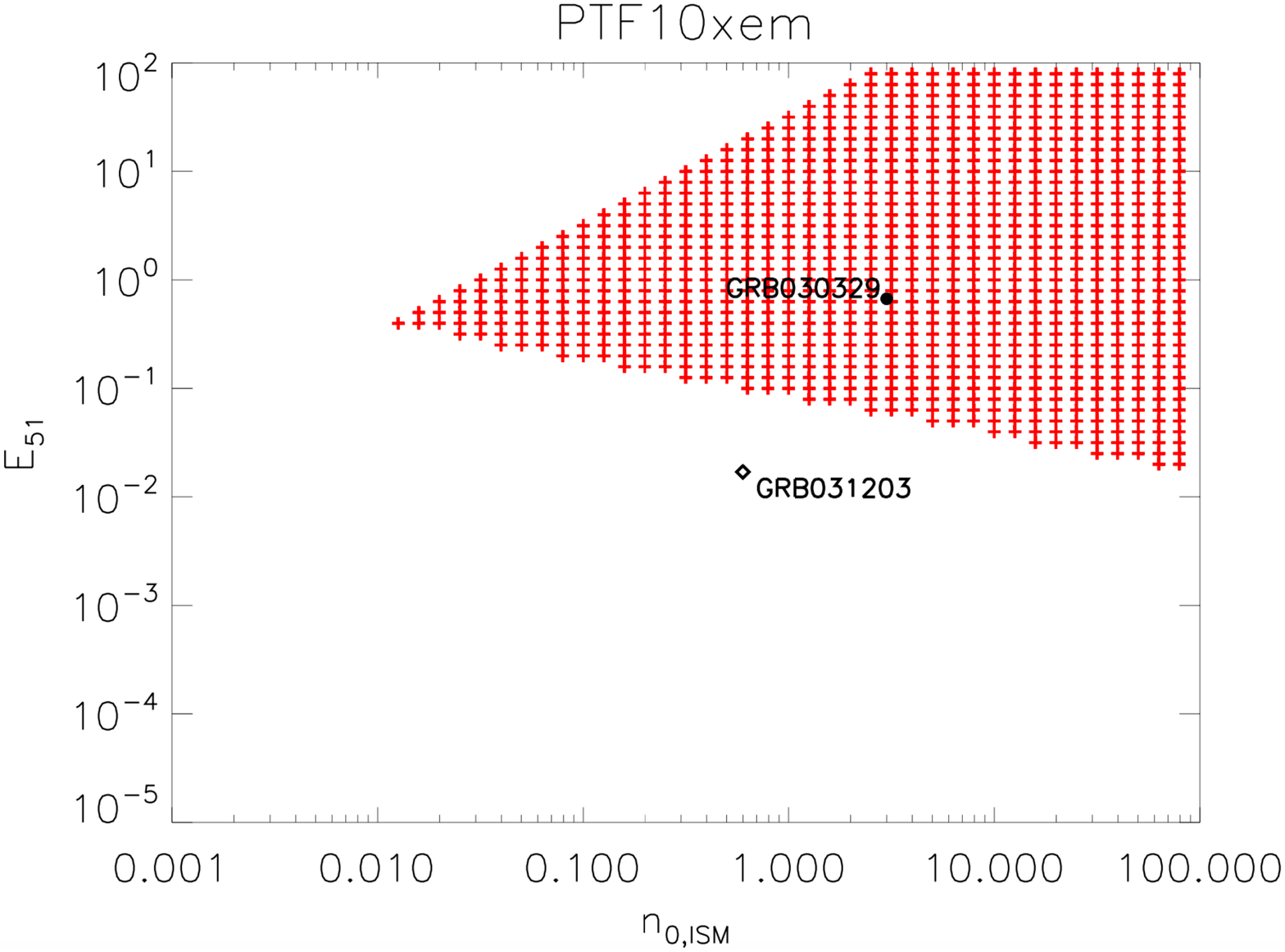}}
\hbox{
\includegraphics[width=5.cm,angle=-90]{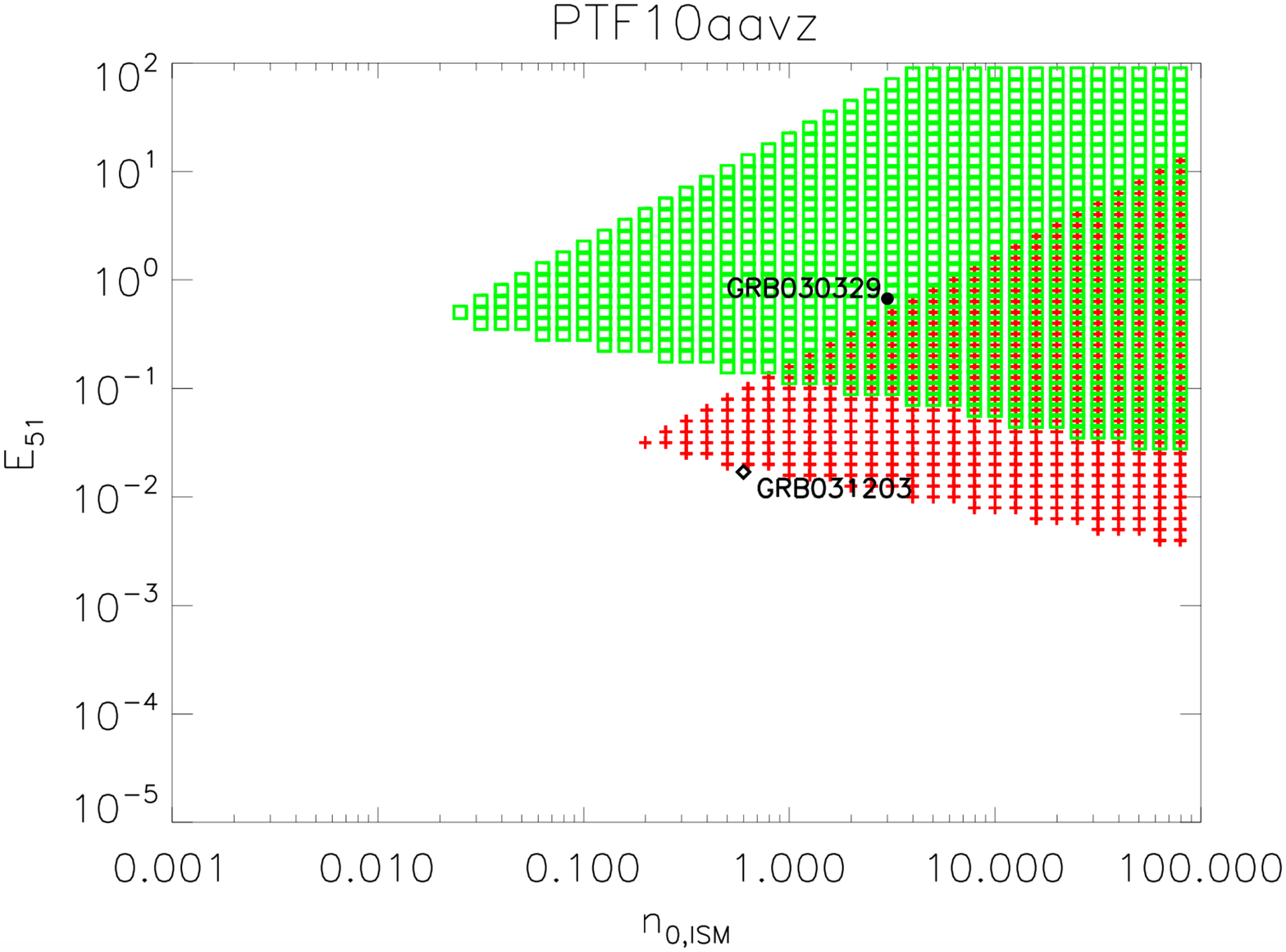}
\hspace{-0.2cm}
\includegraphics[width=5.cm,angle=-90]{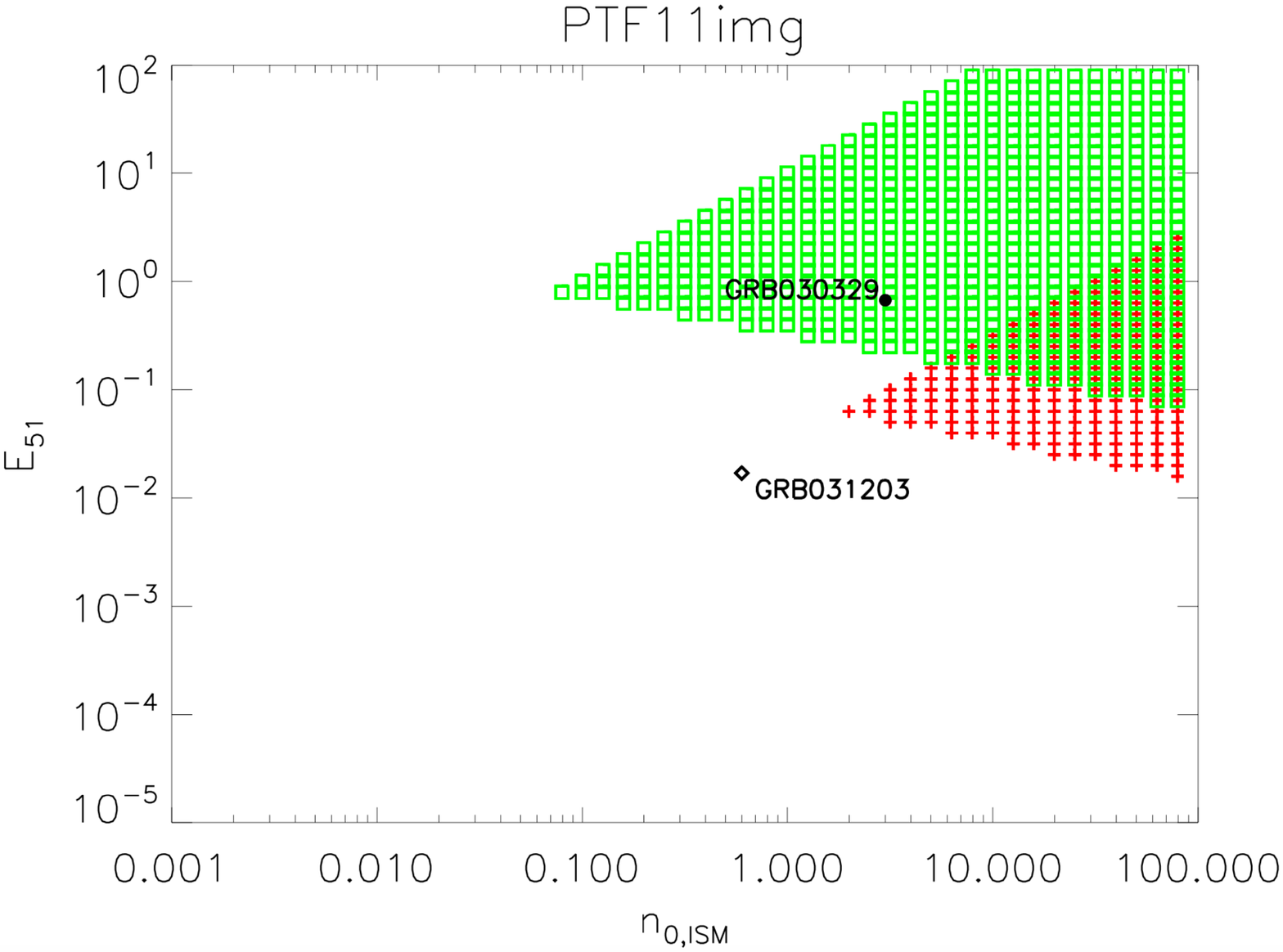}
\hspace{-0.2cm}
\includegraphics[width=5.cm,angle=-90]{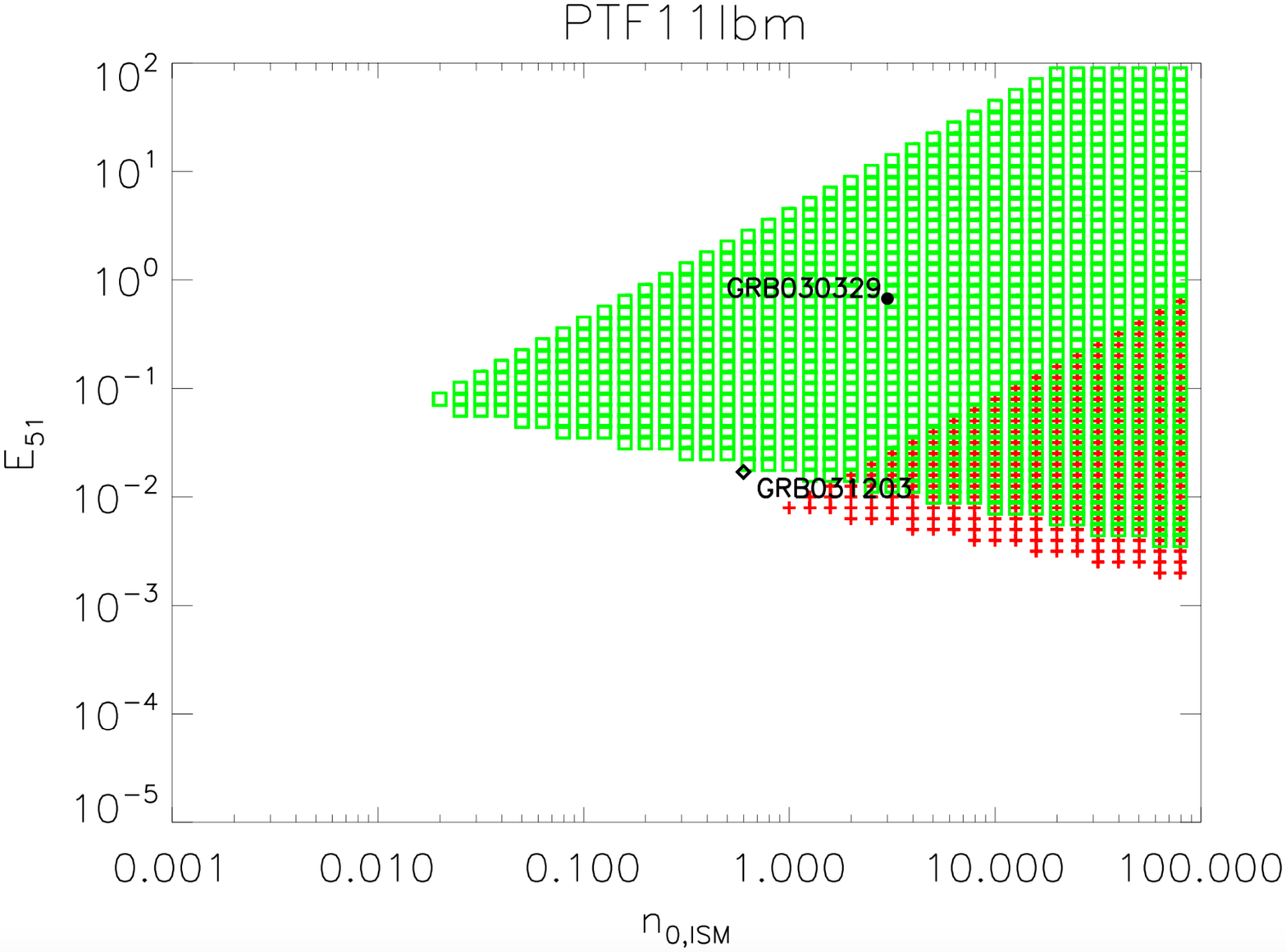}}
\hbox{
\includegraphics[width=5.cm,angle=-90]{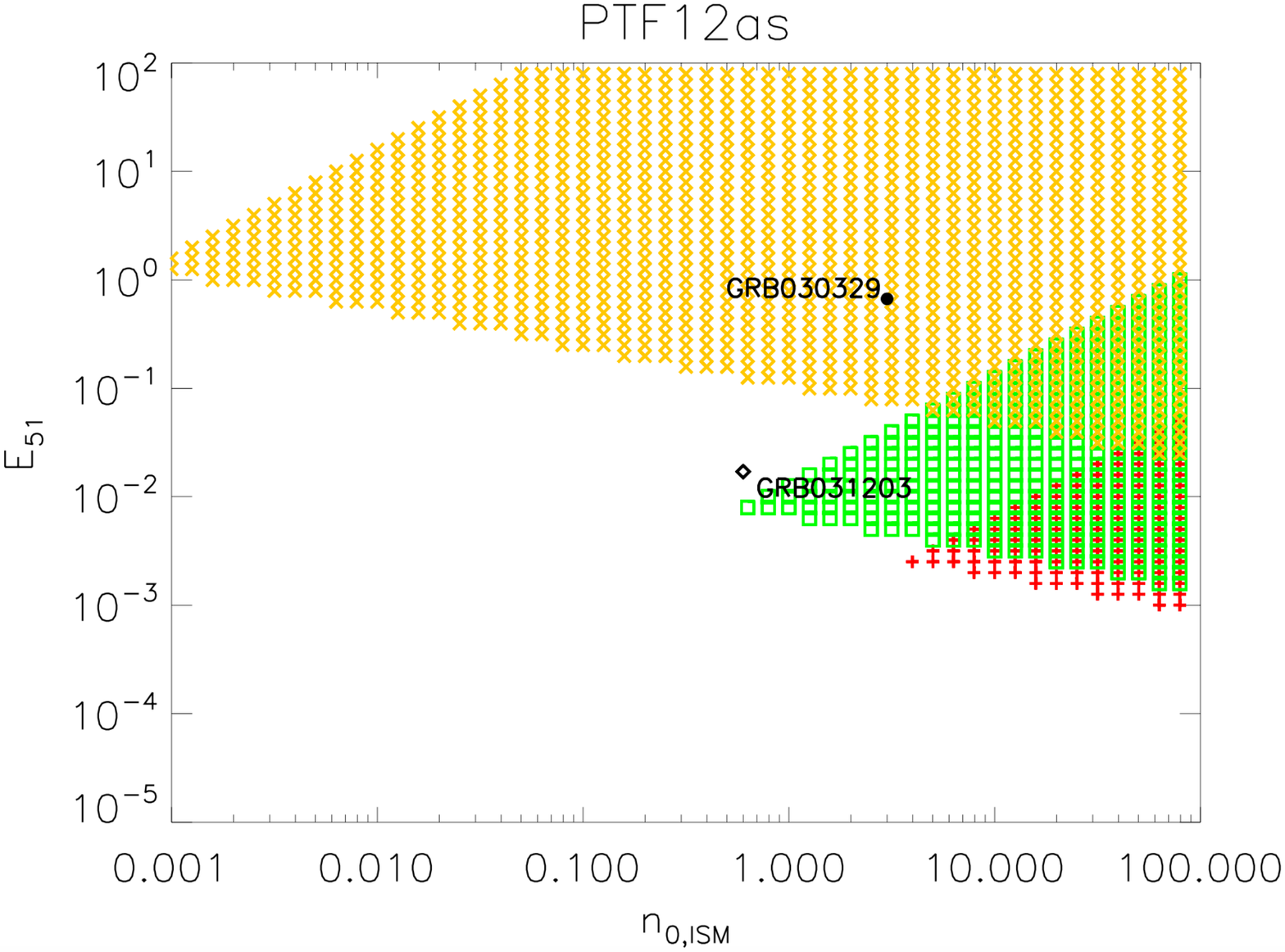}
\hspace{-0.2cm}
\includegraphics[width=5.cm,angle=-90]{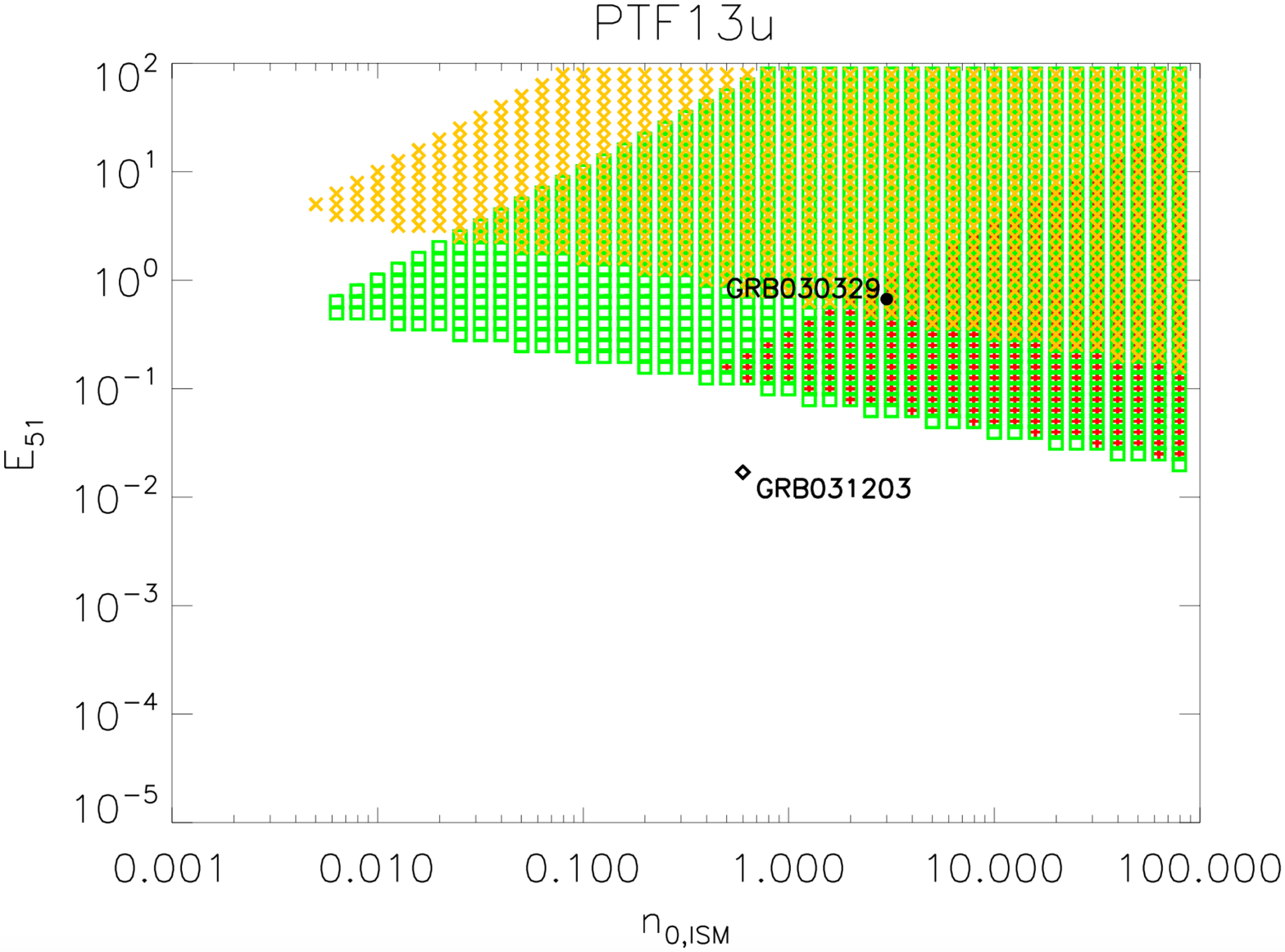}
\hspace{-0.2cm}
\includegraphics[width=5.cm,angle=-90]{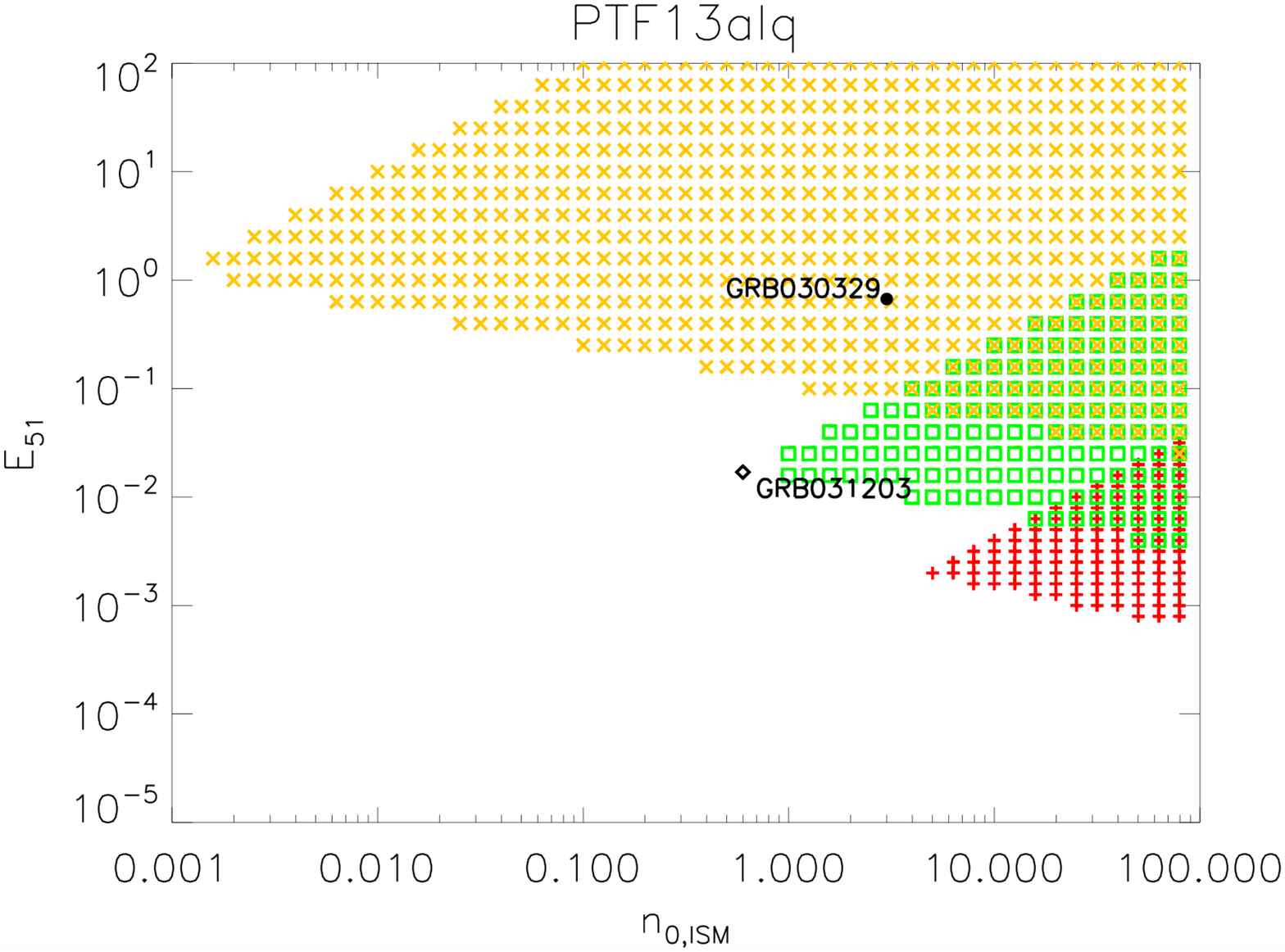}}
\hbox{
\includegraphics[width=5.cm,angle=-90]{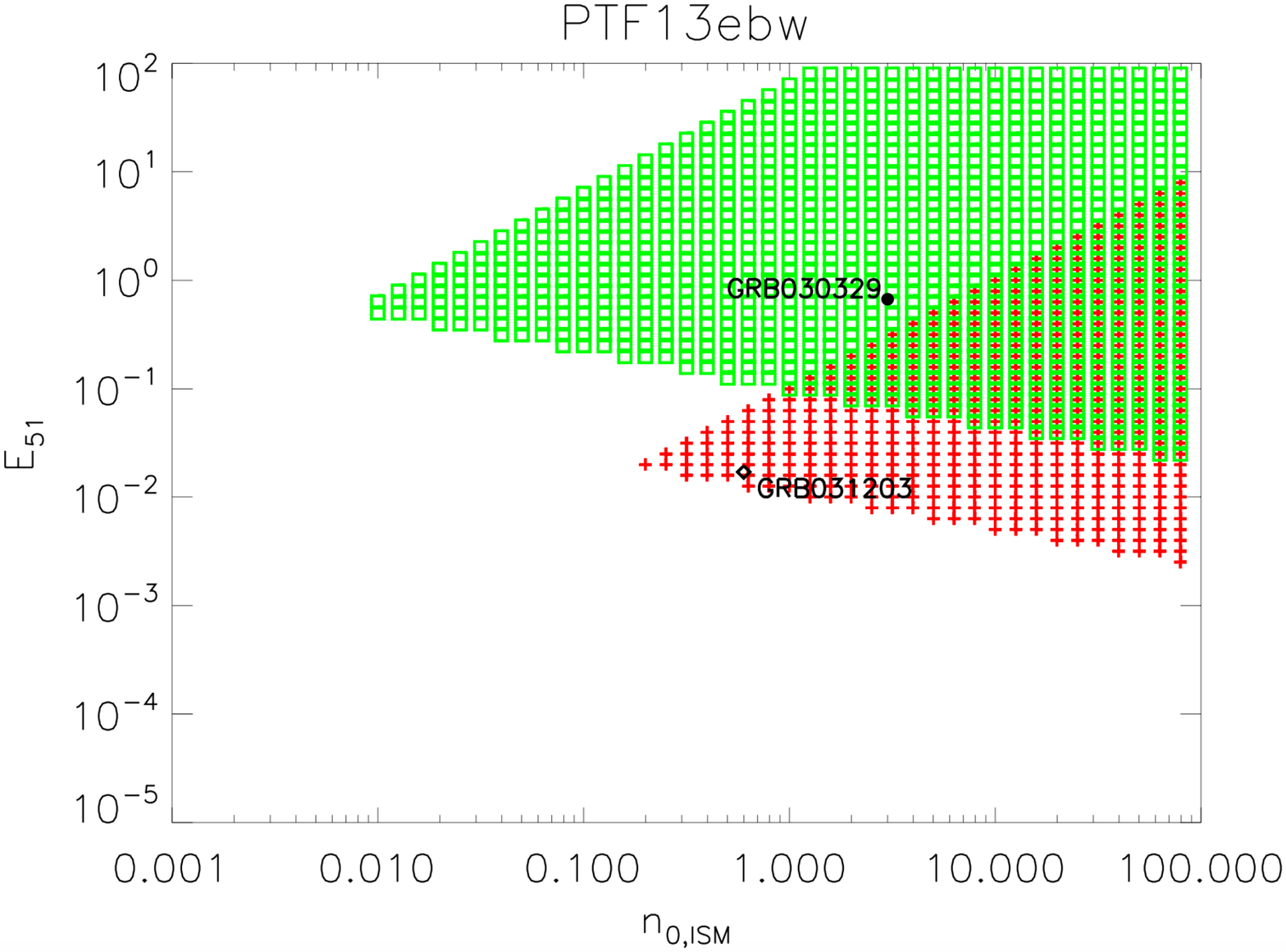}
\hspace{-0.2cm}
\includegraphics[width=5.cm,angle=-90]{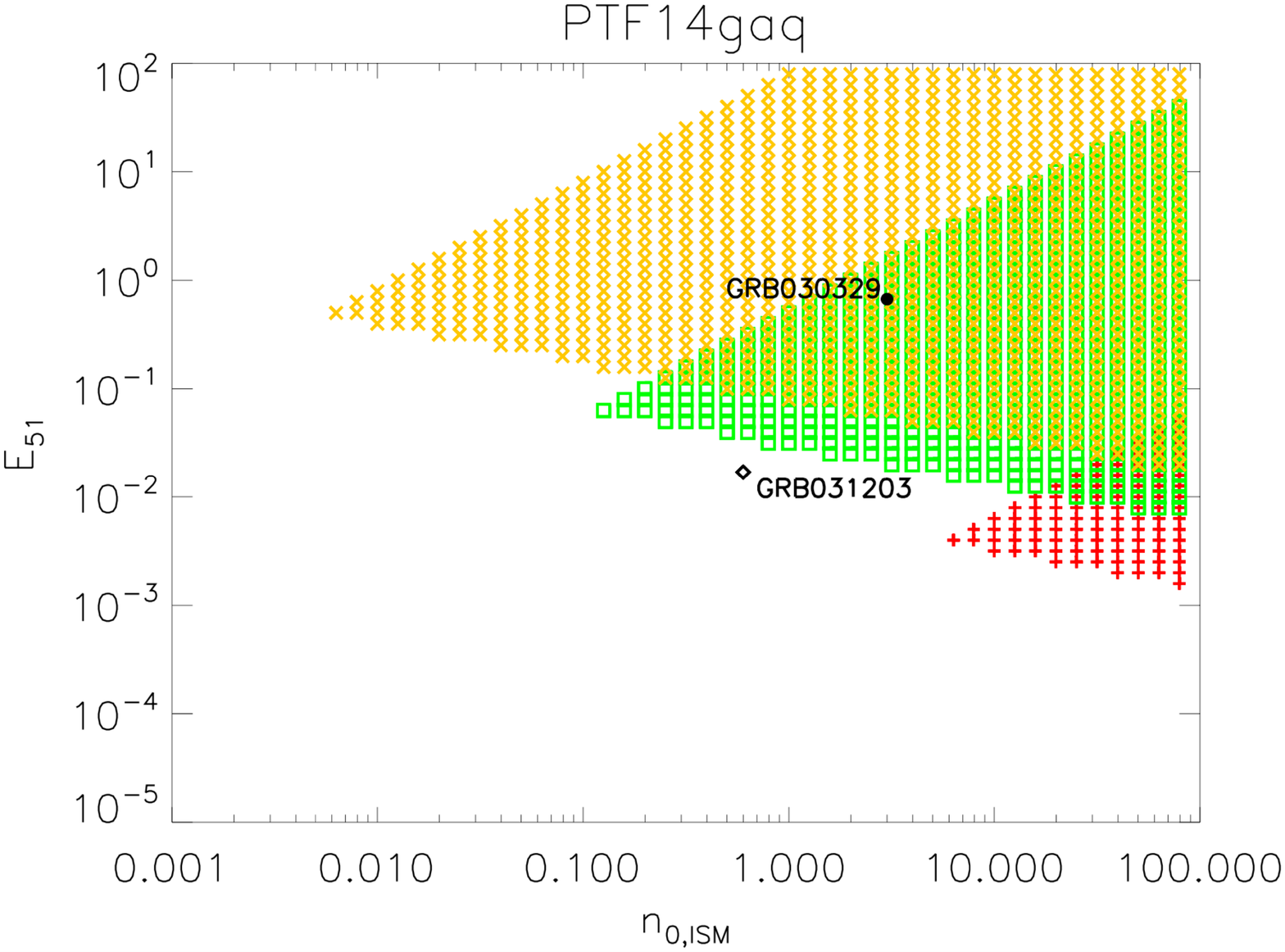}
\hspace{-0.2cm}
\includegraphics[width=5.cm,angle=-90]{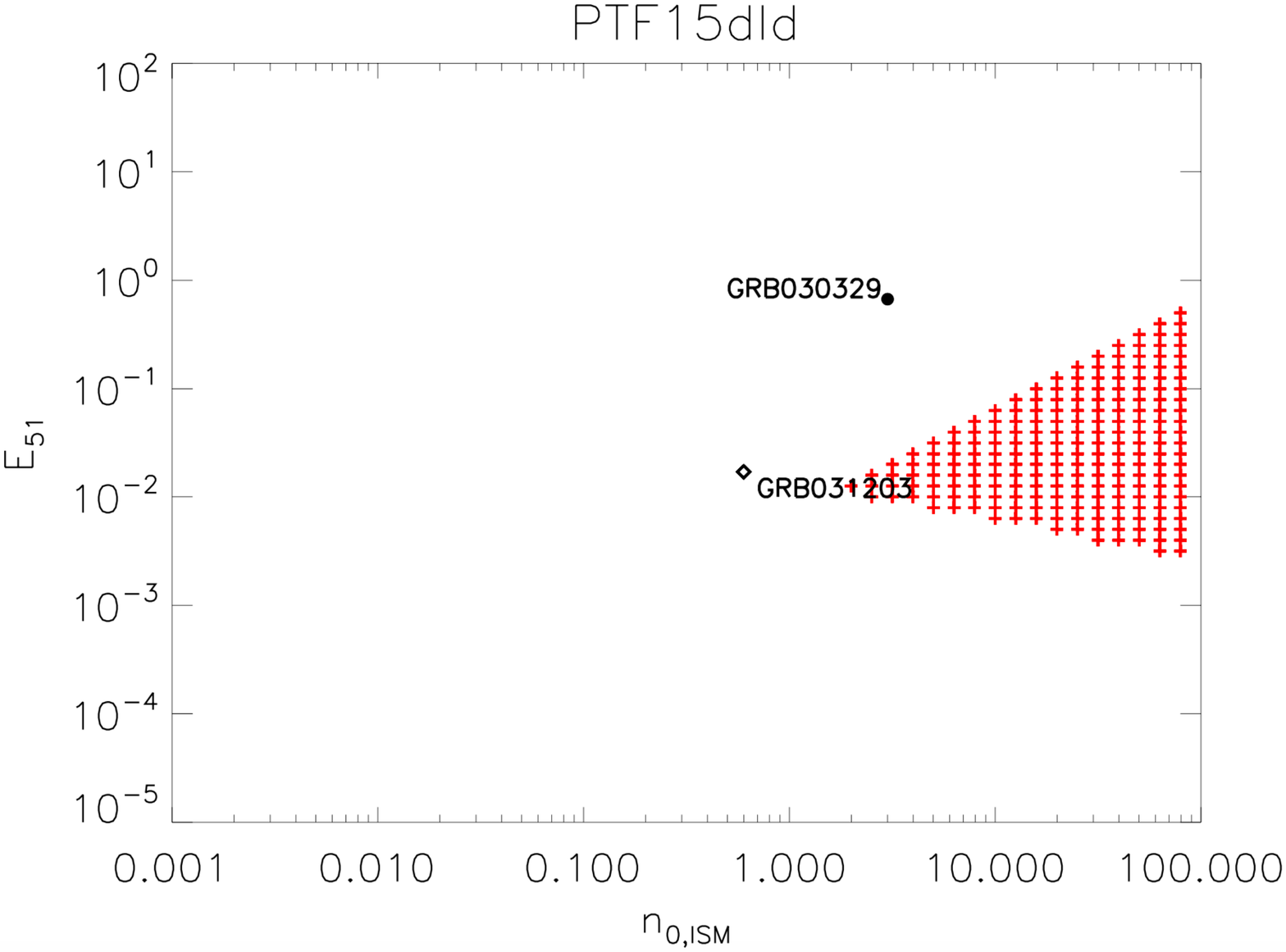}}
\caption{Regions of the energy ($E_{51}$) - density ($n_{0,ISM}$) parameter space excluded by our VLA upper-limits. For each SN in our sample, red, green, and yellow colors correspond to different observations. Specifically, for each SN, we use red for the constraints derived from the first epoch of observations, green for the second epoch (if observed twice; see Table \ref{radioTab}), and yellow for the third epoch (if available; see Table \ref{radioTab}). \label{Fig:radioULISM}}
\end{center}
\end{figure*}

By imposing
 \begin{eqnarray}
 t\gtrsim t_{\rm SNT} ~~~& ~~~F_{\nu} (t) \gtrsim F_{obs,\nu}(t),
 \end{eqnarray}
 we thus calculate, for the SNe in our sample (Table \ref{radioTab}), the values of (beaming corrected) energy and medium density that would give a radio luminosity above our upper-limit $F_{obs,\nu}(t)$, at the time $t$ of our observation. The exclusion regions obtained in this way are shown in Fig. \ref{Fig:radioULISM}. In this Figure we have set the micro-physics parameters equal to their median values as estimated by \citet{Santana2014} i.e., $\epsilon_e\approx 0.22$ and $\epsilon_B\approx 0.01$. However, these parameters (and especially $\epsilon_B$) vary within large ranges, $0.02\lesssim \epsilon_e\lesssim 0.6$ and $3\times10^{-5}\lesssim \epsilon_B \lesssim 0.33$ \citep{Santana2014}. As evident from Eq. (\ref{eqZhang}), for a given upper-limit on the flux at a certain epoch, the smaller the value of the product $\epsilon_e\epsilon_B^{3/4}$, the larger the minimum $E_{51}$ excluded for each value of $n
 _0$ (thus, off-axis emission from lower-energy fireballs / low-luminosity GRBs is less constrained). 
 
\begin{figure*}
\begin{center}
\hbox{
\includegraphics[width=5.cm,angle=-90]{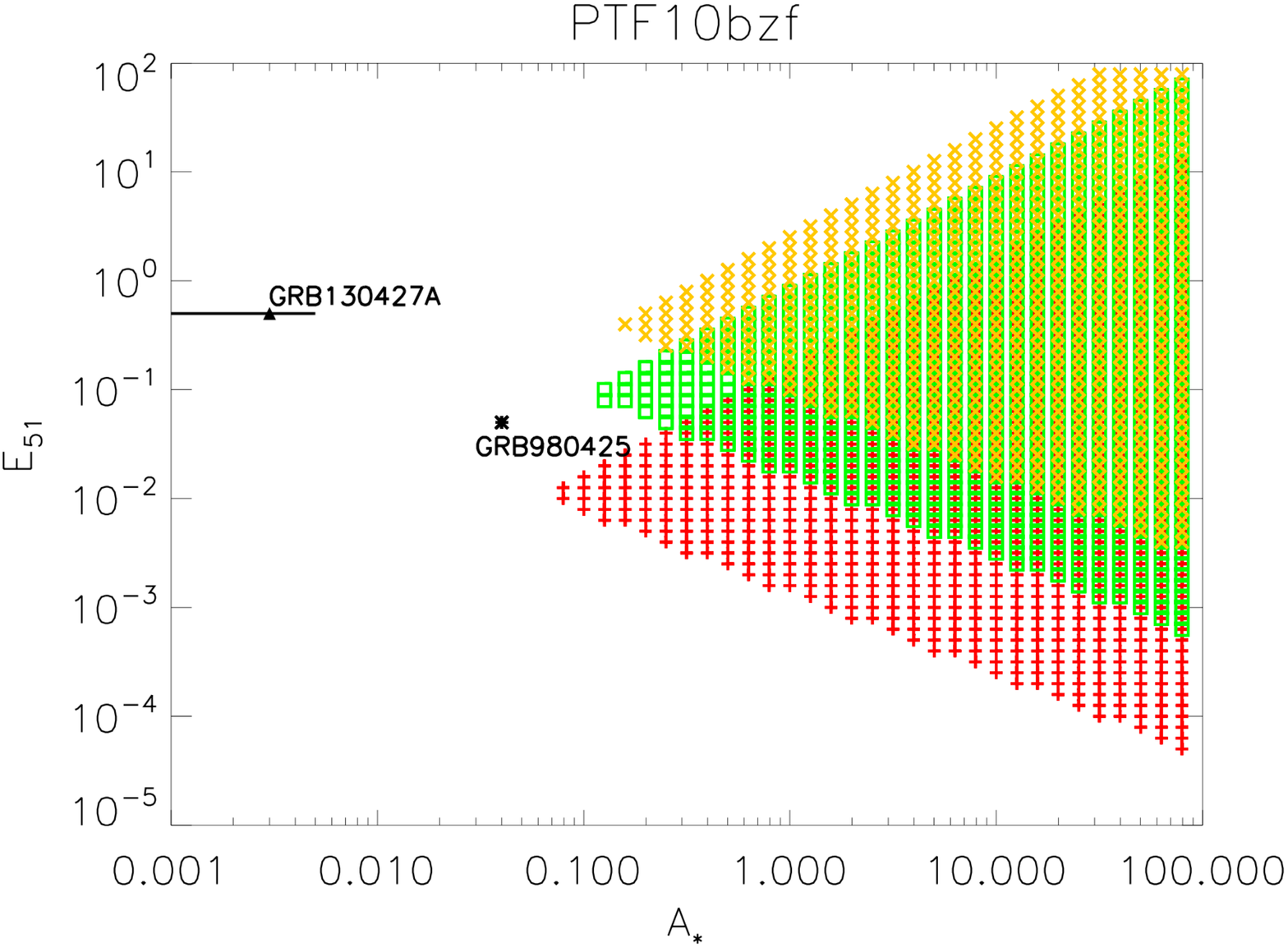}
\hspace{-0.2cm}
\includegraphics[width=5.cm,angle=-90]{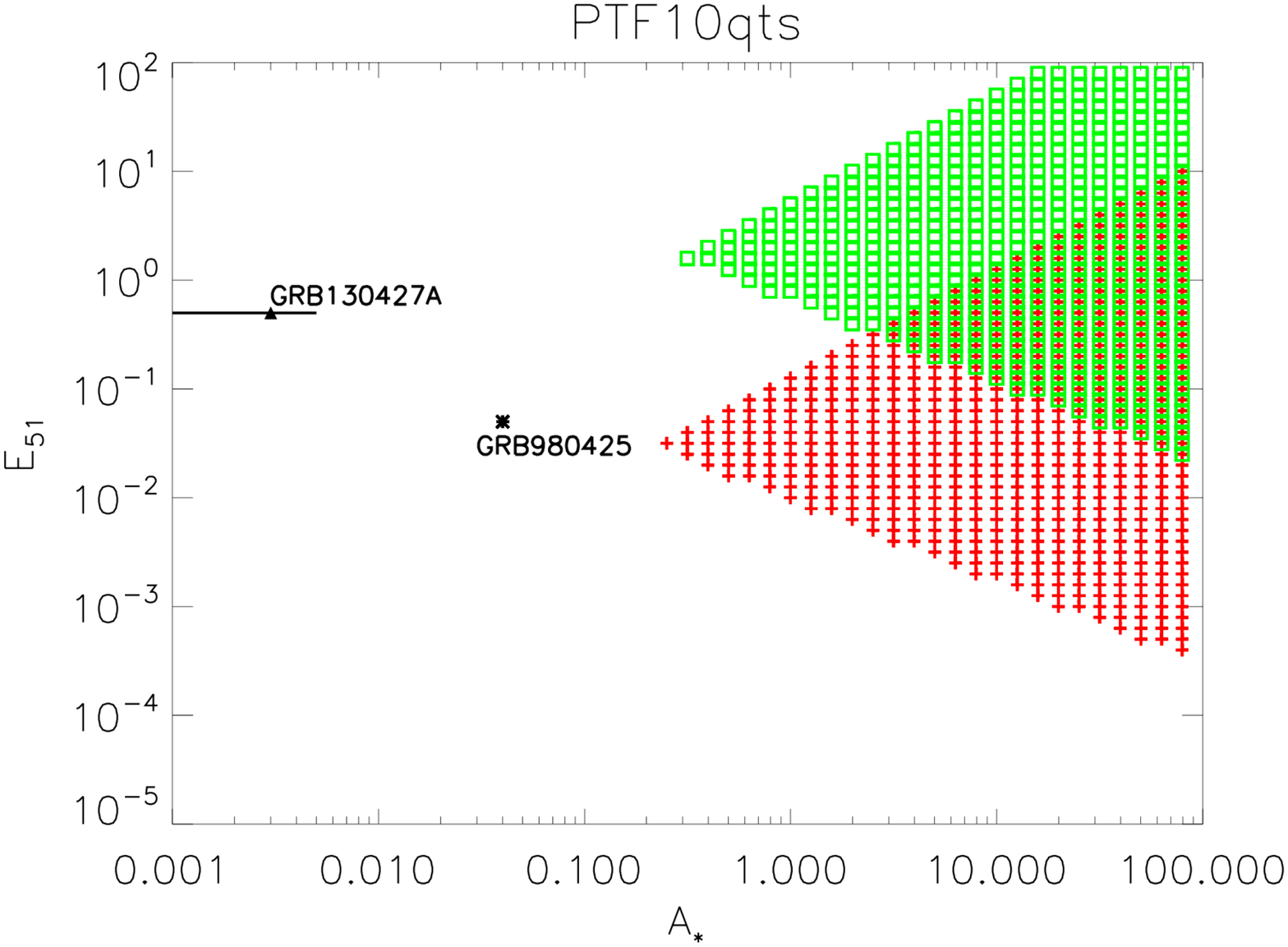}
\hspace{-0.2cm}
\includegraphics[width=5.cm,angle=-90]{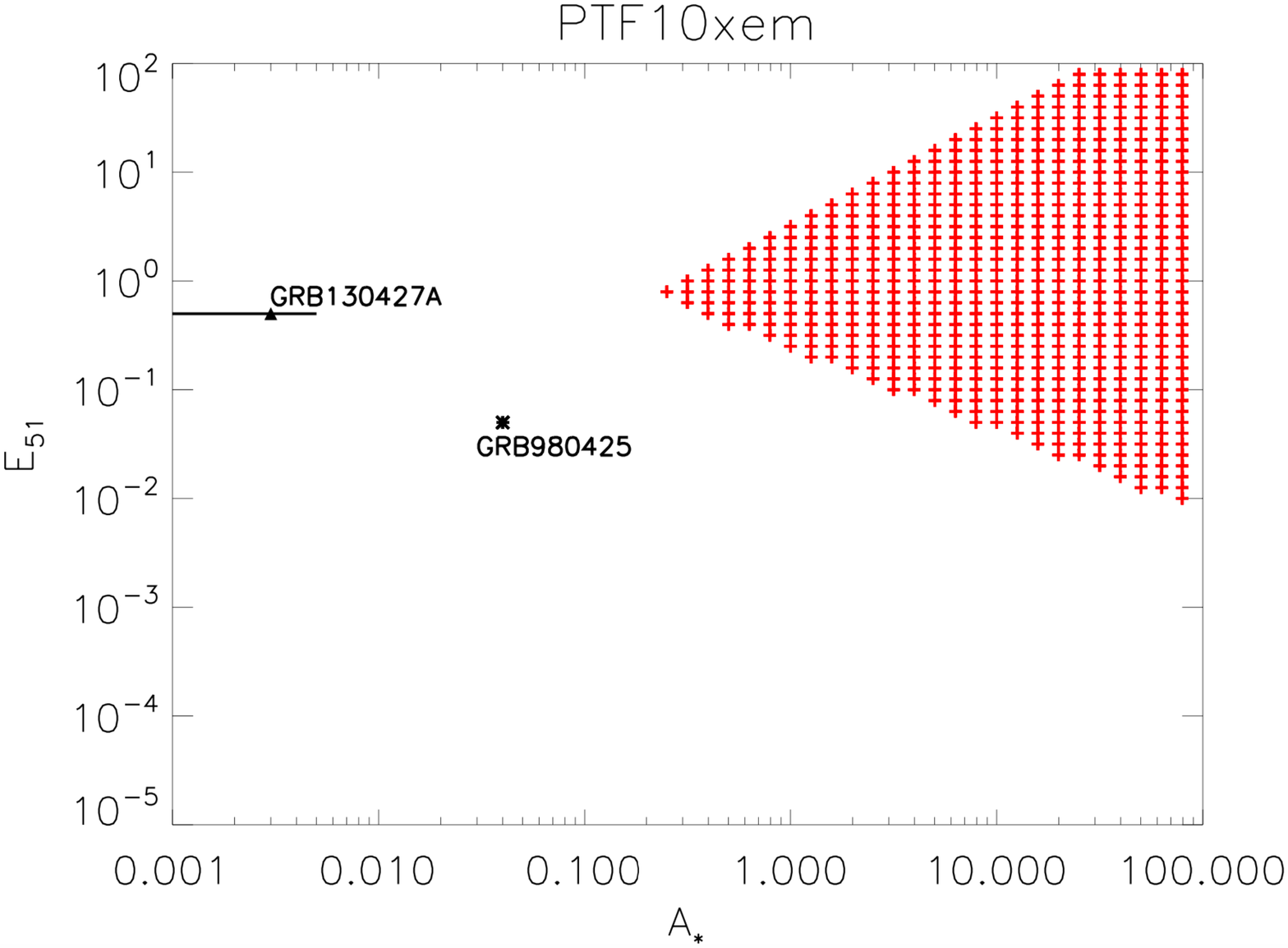}}
\hbox{
\includegraphics[width=5.cm,angle=-90]{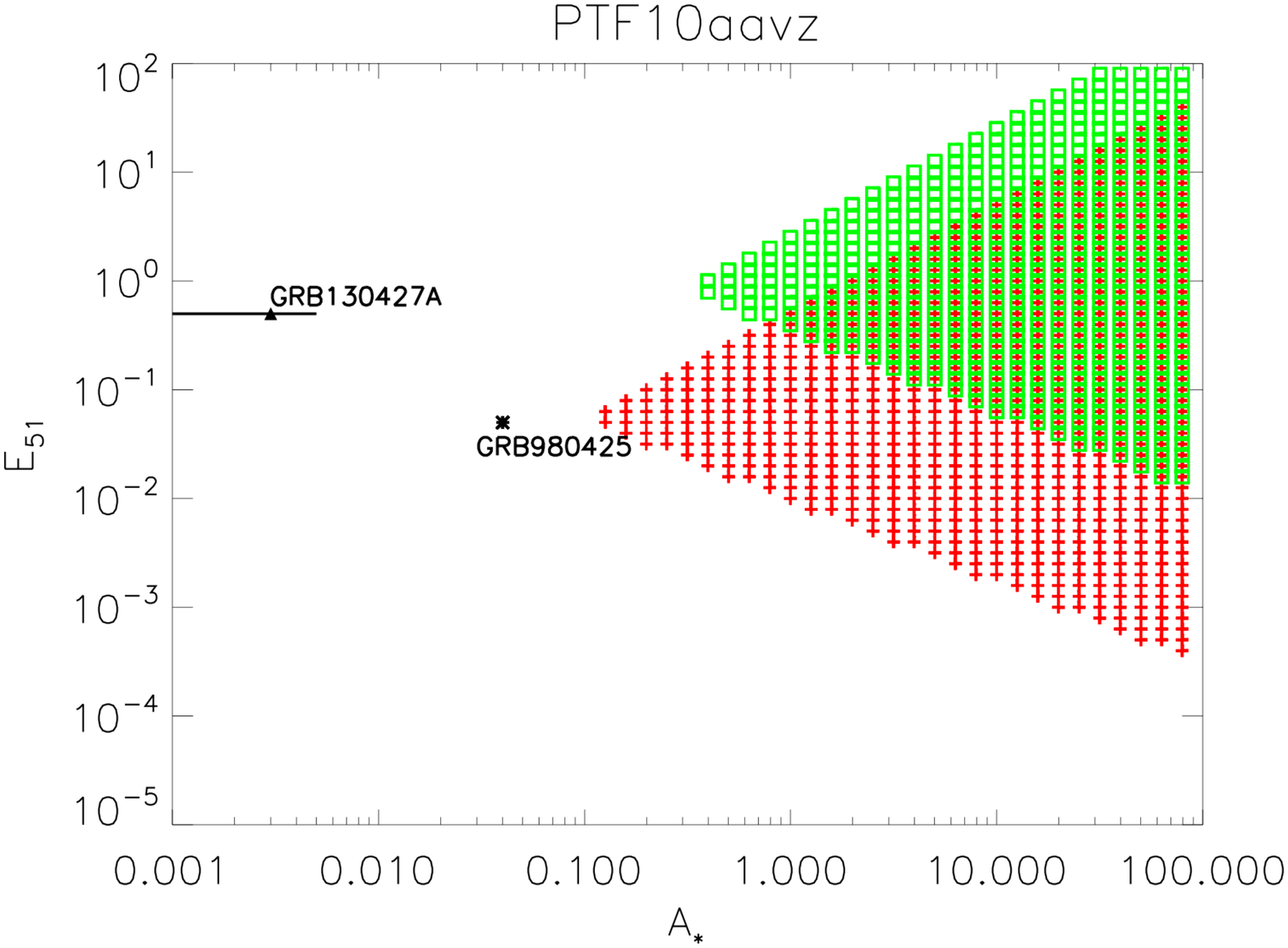}
\hspace{-0.2cm}
\includegraphics[width=5.cm,angle=-90]{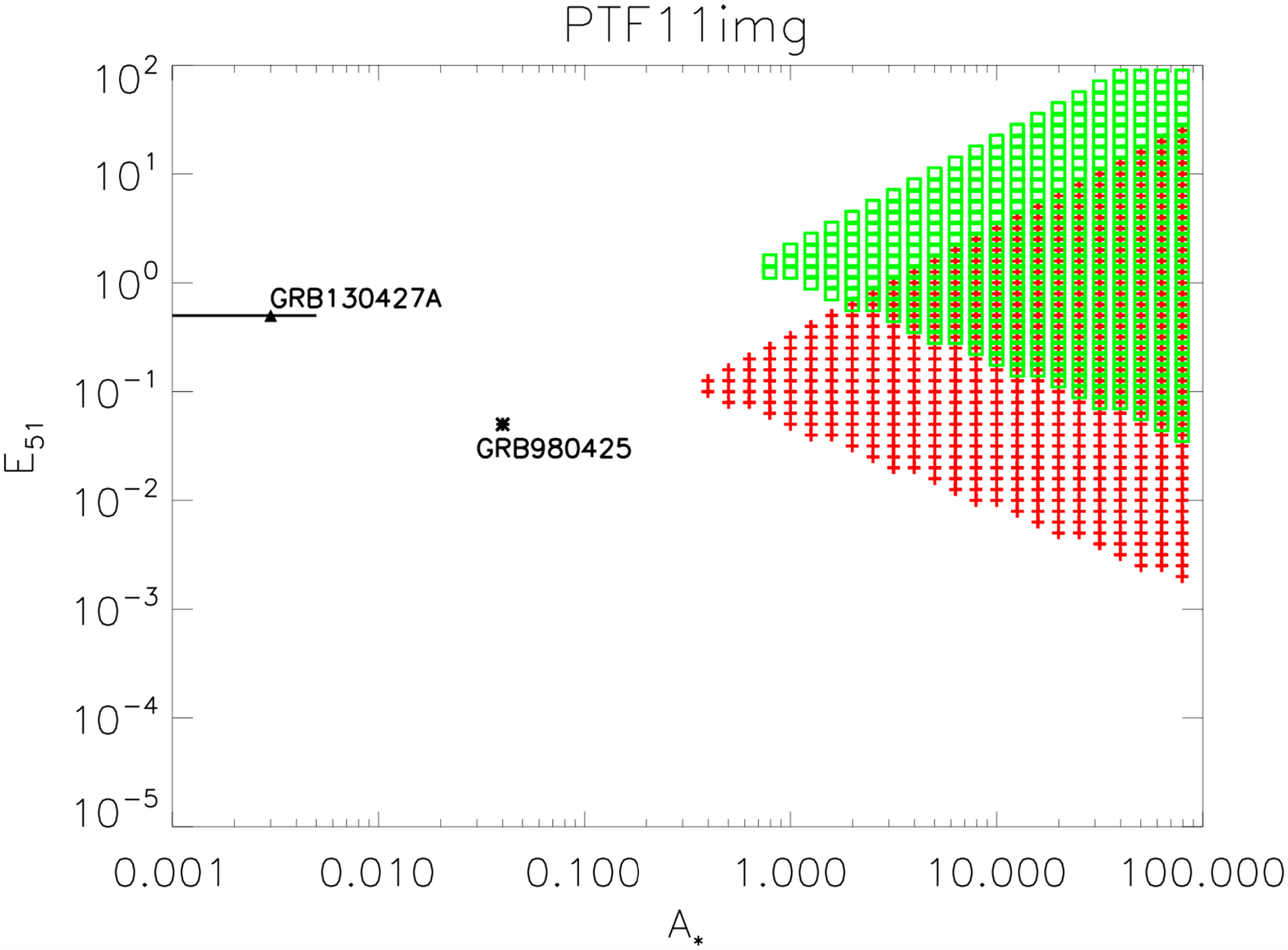}
\hspace{-0.2cm}
\includegraphics[width=5.cm,angle=-90]{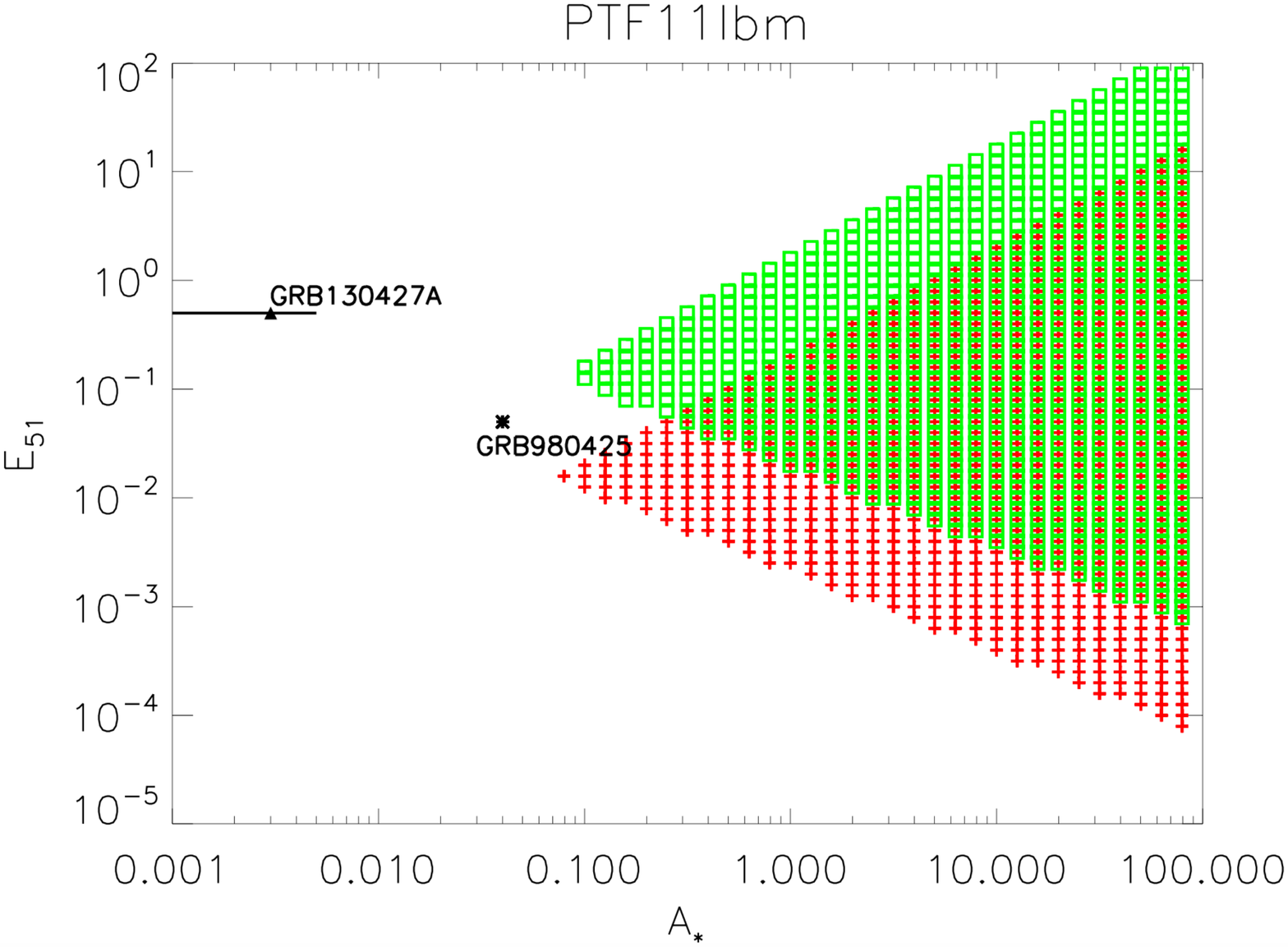}}
\hbox{
\includegraphics[width=5.cm,angle=-90]{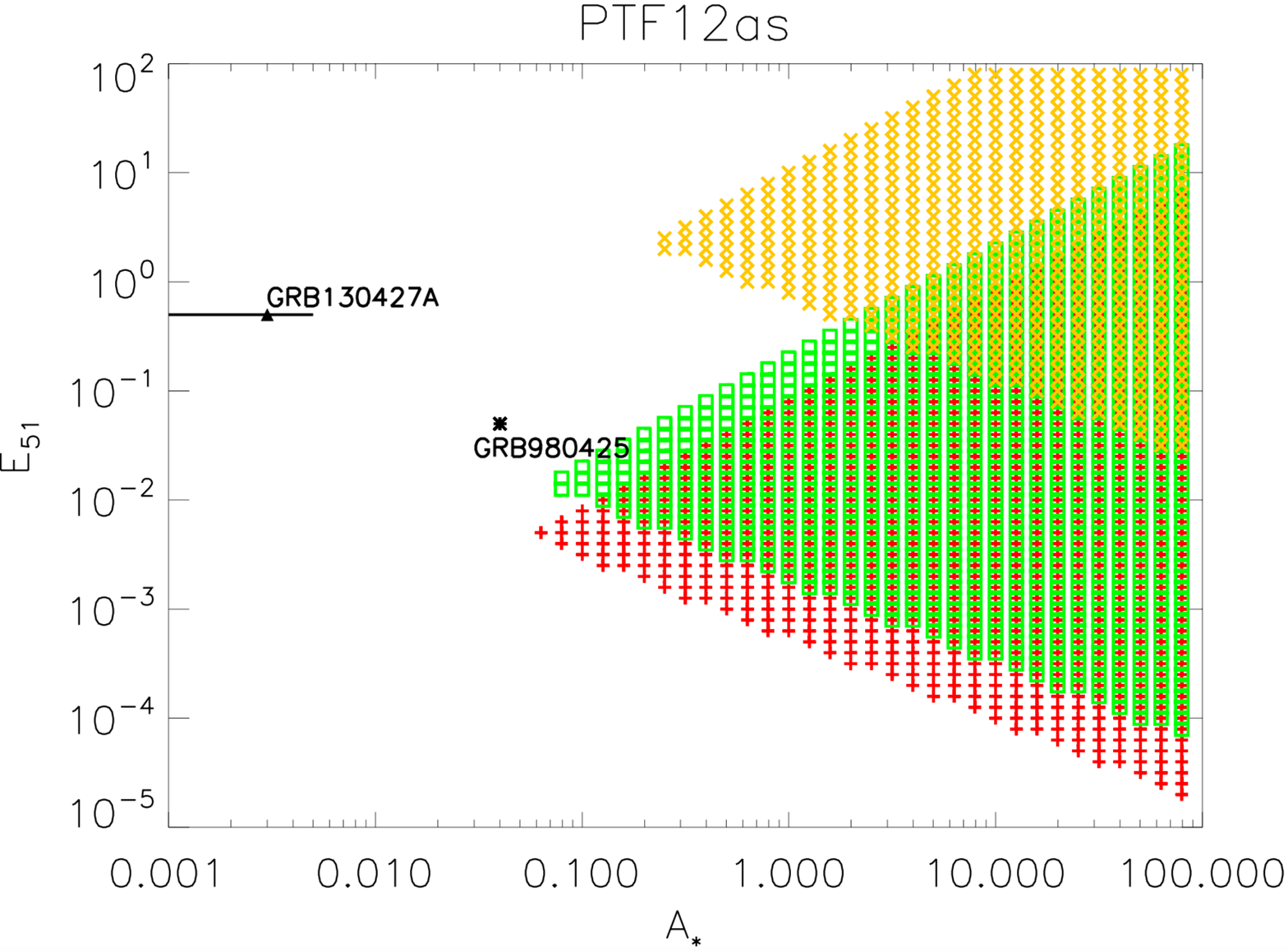}
\hspace{-0.2cm}
\includegraphics[width=5.cm,angle=-90]{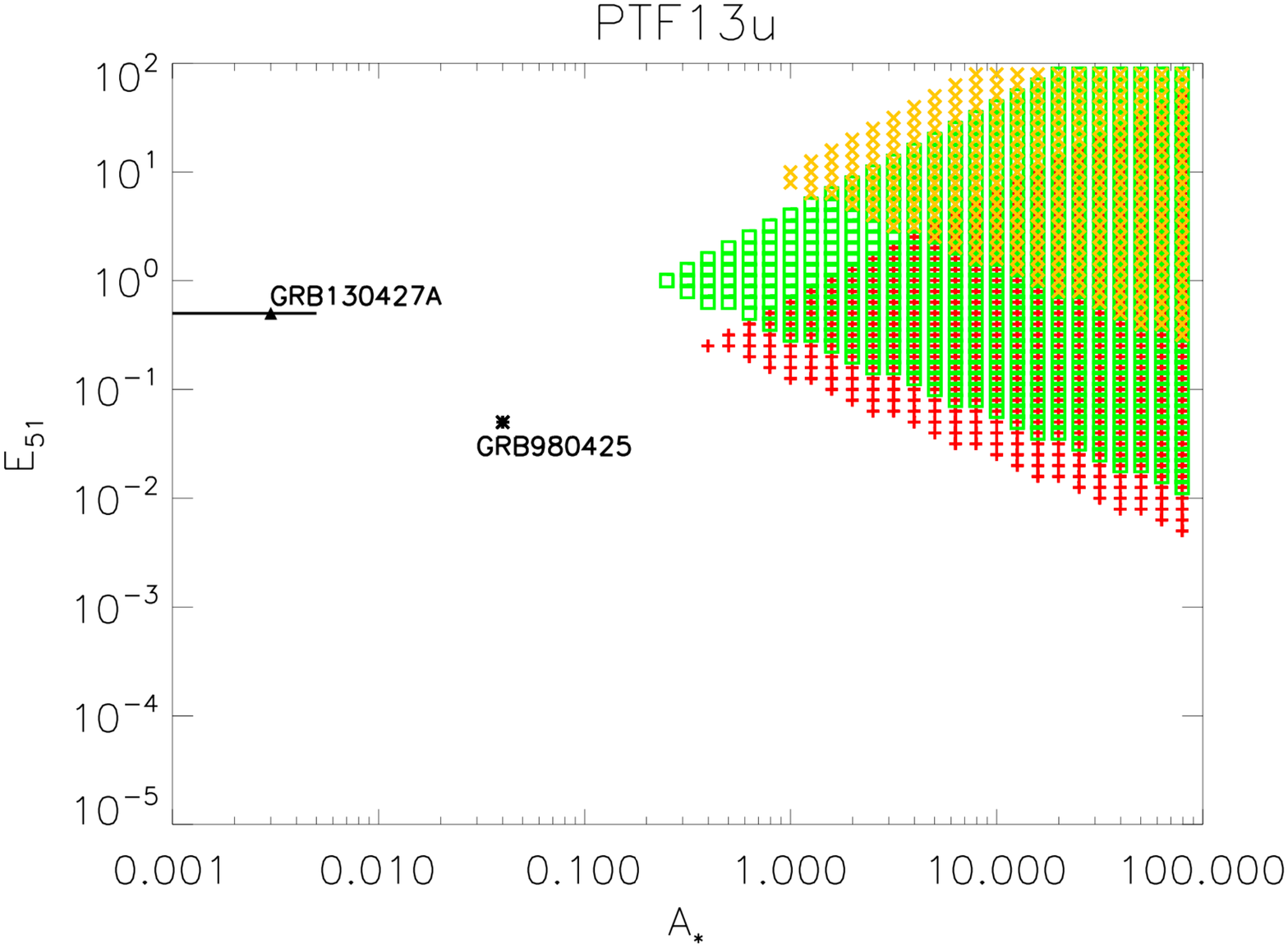}
\hspace{-0.2cm}
\includegraphics[width=5.cm,angle=-90]{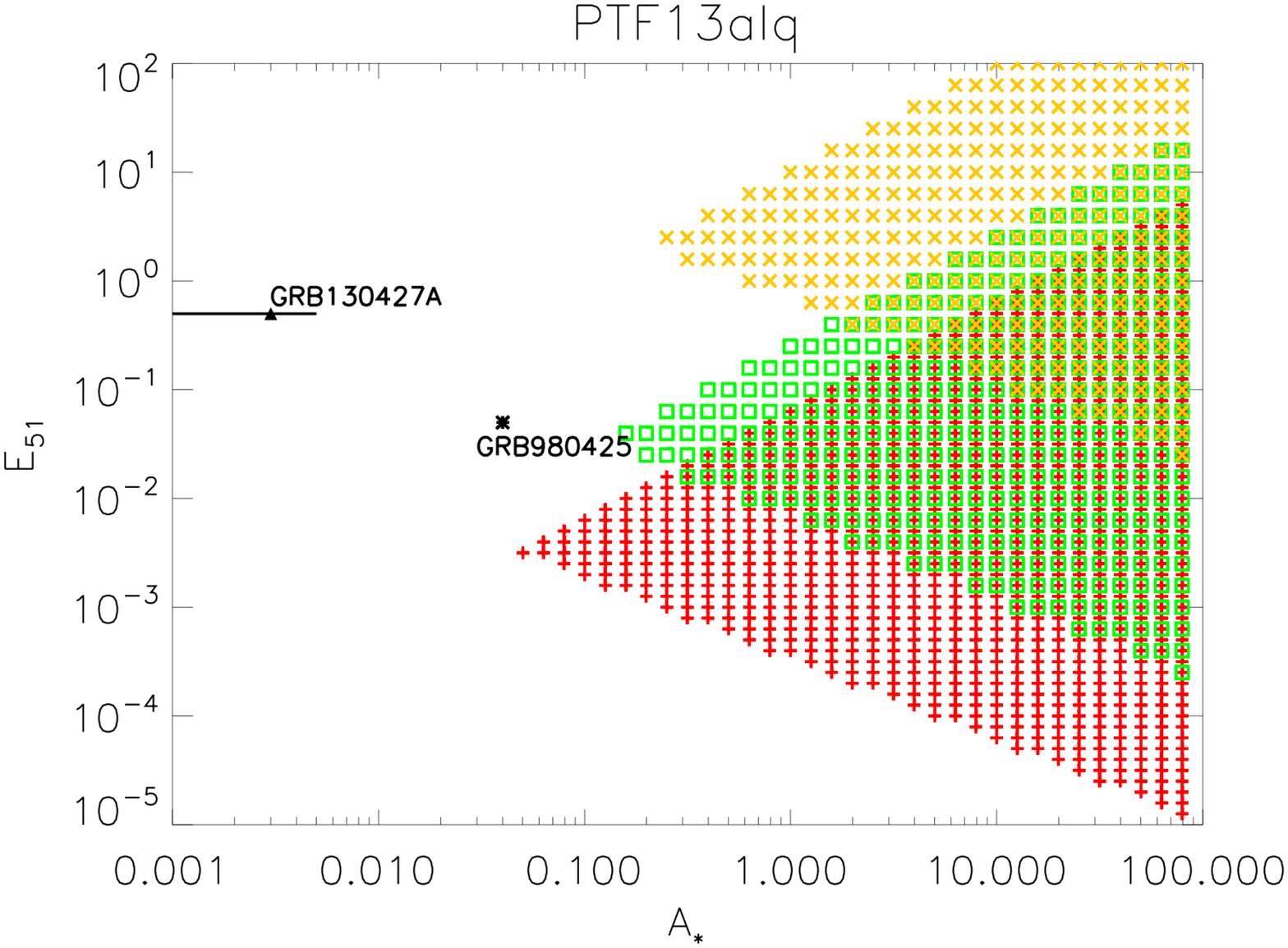}}
\hbox{
\includegraphics[width=5.cm,angle=-90]{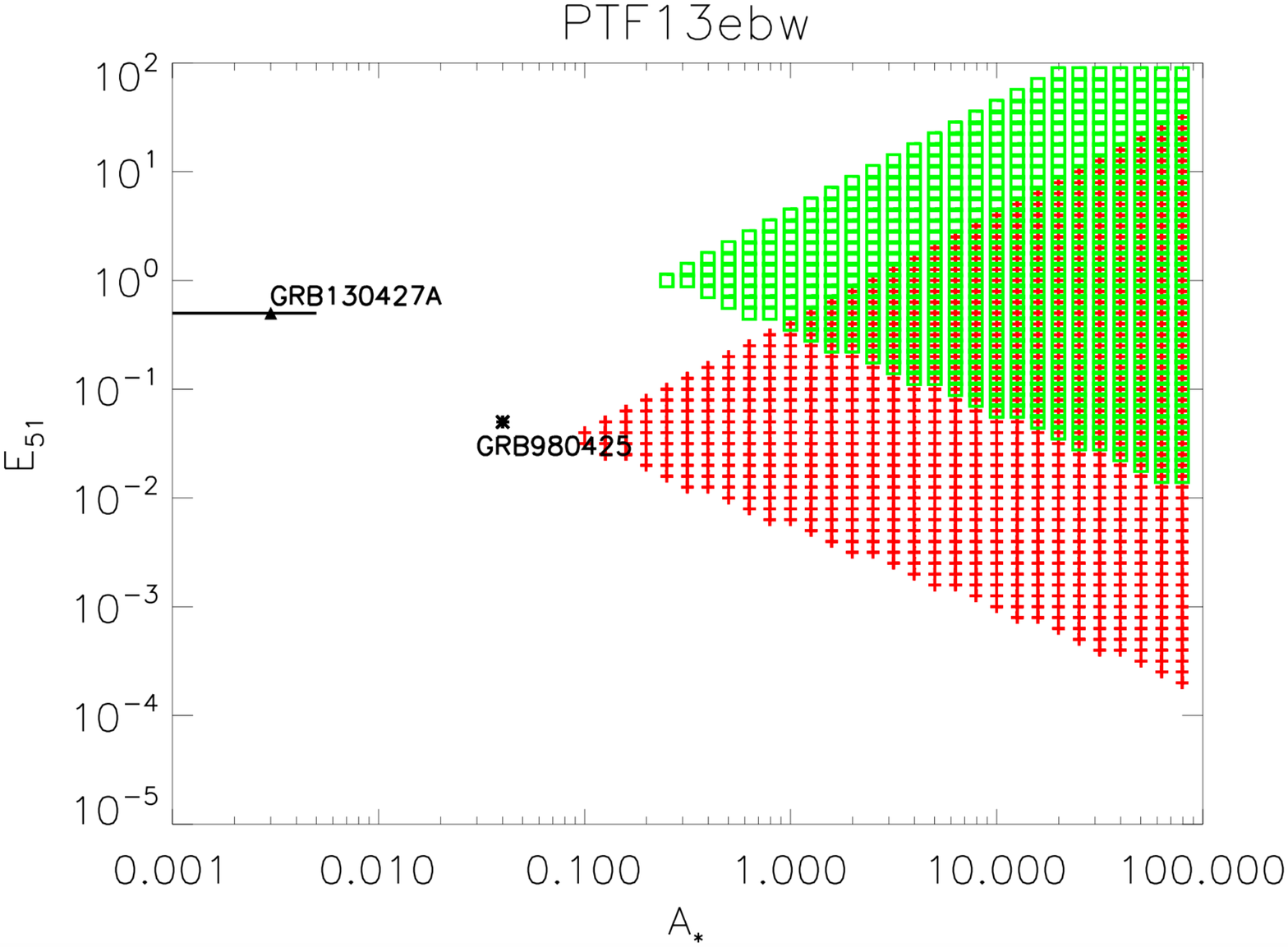}
\hspace{-0.2cm}
\includegraphics[width=5.cm,angle=-90]{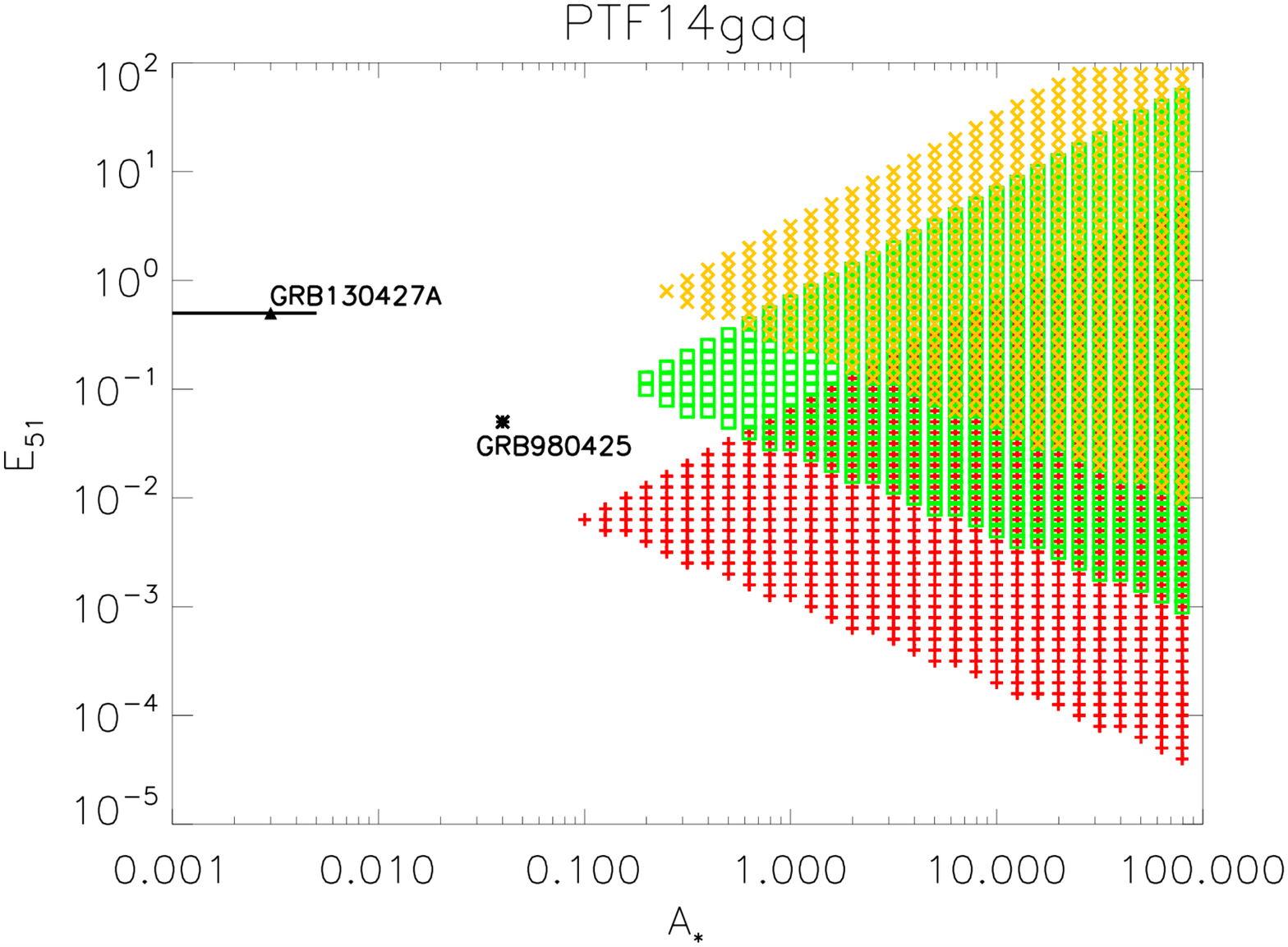}
\hspace{-0.2cm}
\includegraphics[width=5.cm,angle=-90]{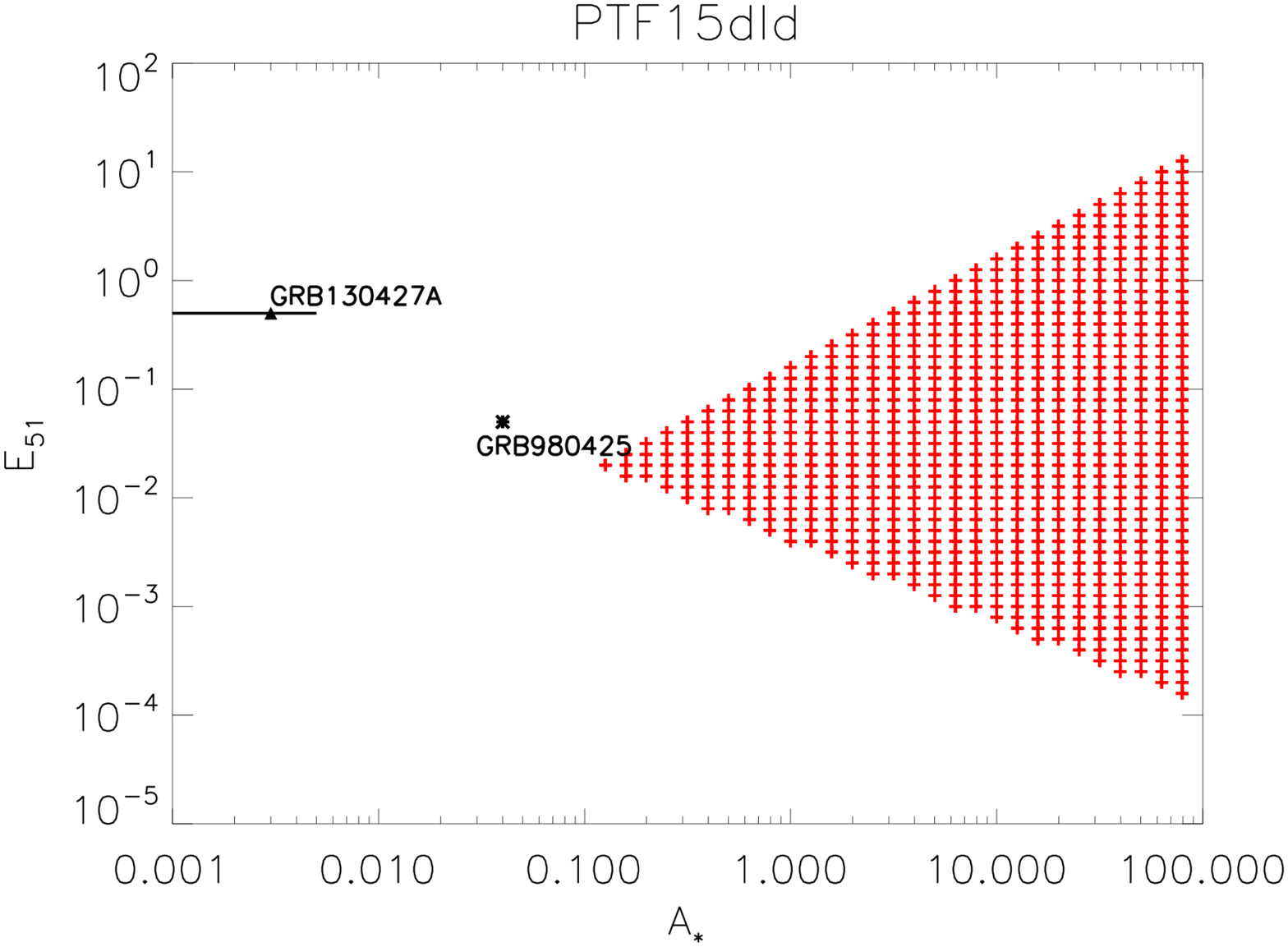}}
\caption{Regions of the energy ($E_{51}$) - density ($A_*$) parameter space excluded by our VLA upper-limits. For each SN in our sample, red, green, and yellow colors correspond to different observations. Specifically, for each SN, we use red for the constraints derived from the first epoch of observations, green for the second epoch (if observed twice; see Table \ref{radioTab}), and yellow for the third epoch (if available; see Table \ref{radioTab}). \label{Fig:radioULWind}}
\end{center}
\end{figure*}

 In Fig. \ref{Fig:radioULWind} we use our upper-limits to set similar constraints on off-axis GRBs expanding in a wind medium. In this case, the expected radio flux is approximated as \citep{Waxman2004,Soderberg2006}:
\begin{eqnarray}
\nonumber F_{\rm \nu}(t)\approx 0.053  \,d^{-2}_{L,28}(1+z)^{1/2}\left(\frac{\epsilon_{\rm e}}{0.1}\right)\left(\frac{\epsilon_{\rm B}}{0.1}\right)^{3/4}A^{3/4}_{*}E_{51}\\ \times\left(\frac{\nu}{1\,{\rm GHz}}\right)^{-1/2}\left(\frac{t}{92\,{\rm d}(1+z)}\right)^{-3/2}{\rm\,mJy},~~~
\label{eqlate}
\end{eqnarray}
where $A_*$ is the circumstellar density, which is related to progenitor mass loss-rate $\dot{M}$ and wind velocity ${\rm v_w}$ as $A_*=(\dot{M}/10^{-5}M_{\odot}{\rm yr}^{-1})/({\rm v_w}/1000\,{\rm km\,s}^{-1})$; and where we have used \citep{Waxman2004,Soderberg2006}:
\begin{equation}
t_{\rm SNT} \approx 92\,{\rm d} \left(\frac{E_{51}}{A_*}\right).
\label{tnr}
\end{equation}
We note that Eq. (\ref{eqlate}) corresponds to  Eq. (14) in \citet{Waxman2004} where the dependence on $t_{\rm SNT}$ is eliminated by using our Eq. \ref{tnr} \citep[or, equivalently, Eq. (8) in ][]{Waxman2004}.  Here we are assuming that the conclusions reached by \citet{Zhang2009} for the constant density case are valid also for a fireball expanding in a wind medium, namely, that Eq. (\ref{eqlate}) provides an estimate of the fireball parameters good to within a factor of $\approx 2$ when compared to flux measurements carried out at epochs $t\gtrsim t_{\rm SNT}$.

We can compare the results shown in Figs. 6-7 with the energy and density derived from the broad-band afterglow modeling of the high-luminosity GRB\,030329 \citep[for which $\theta_j\approx 5-17$\,deg, $E_{51}=0.67$, $n_{0,ISM}\approx 3$;][]{Berger2003b,Soderberg2006} and GRB\,130427A  \citep[for which $E_{51}\gtrsim 0.5$, $\theta_j\gtrsim 5$\,deg, and $0.01\lesssim A_*\lesssim 0.05$;][]{Perley2014}, and of the low-luminosity GRB\,980425 \citep[for which $E_{51}\approx 0.05$ and $A_*\approx 0.04$;][]{Waxman2004b,Waxman2004,Soderberg2006} and GRB\,031203 \citep[for which $E_{51}=0.017$ and $n_{0,ISM}=0.6$;][]{Soderberg2004,Ramirez2005,Soderberg2006}. From such a comparison we conclude that most of the SNe in our sample exclude GRBs with (beaming corrected) energy and ISM density comparable to GRB\,030329, observed largely off-axis and/or during the non-relativistic phase. On the other hand, our upper-limits cannot exclude a GRB as sub-energetic as GRB\,980425 observed during its non-relativistic phase, nor an off-axis GRB expanding in a low-density environment such as GRB\,031203 and GRB\,130427A.

As discussed before, our conclusions depend somewhat on the assumed values of the micro-physics parameters. Indeed, as evident from Eq. (\ref{eqlate}), also in a wind environment the smaller the value of the product $\epsilon_e\epsilon_B^{3/4}$, the larger the minimum $E_{51}$ excluded for each value of $A_*$. 

\section{VLA detections: PTF11cmh and PTF14dby}
\label{radiomodel}

\begin{figure*}
\begin{center}
\includegraphics[width=12cm,angle=-90]{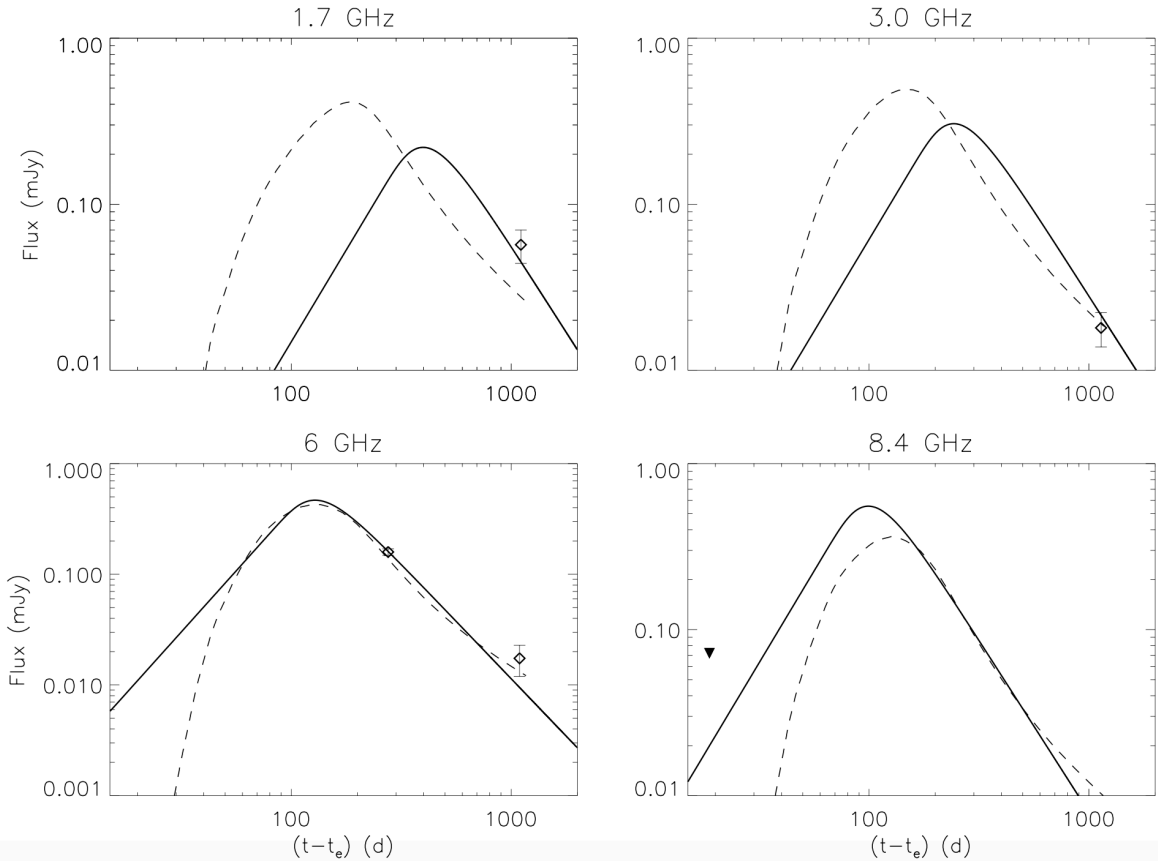}
\caption{Best fit radio light curves of PTF11cmh in the synchrotron self-absorbed radio SN model (solid; see Section \ref{nonrelradio}), and an off-axis GRB model light curve for a fireball expanding in a constant density ISM (dashed; see Section \ref{vaneertenmodels}), compared with our VLA observations (see Table \ref{radioTab}). See text for a discussion of the models and best fit parameters. \label{Fig:radiofits}}
\end{center}
\end{figure*}

\begin{figure*}
\begin{center}
\hbox{
\includegraphics[width=7.1cm,angle=-90]{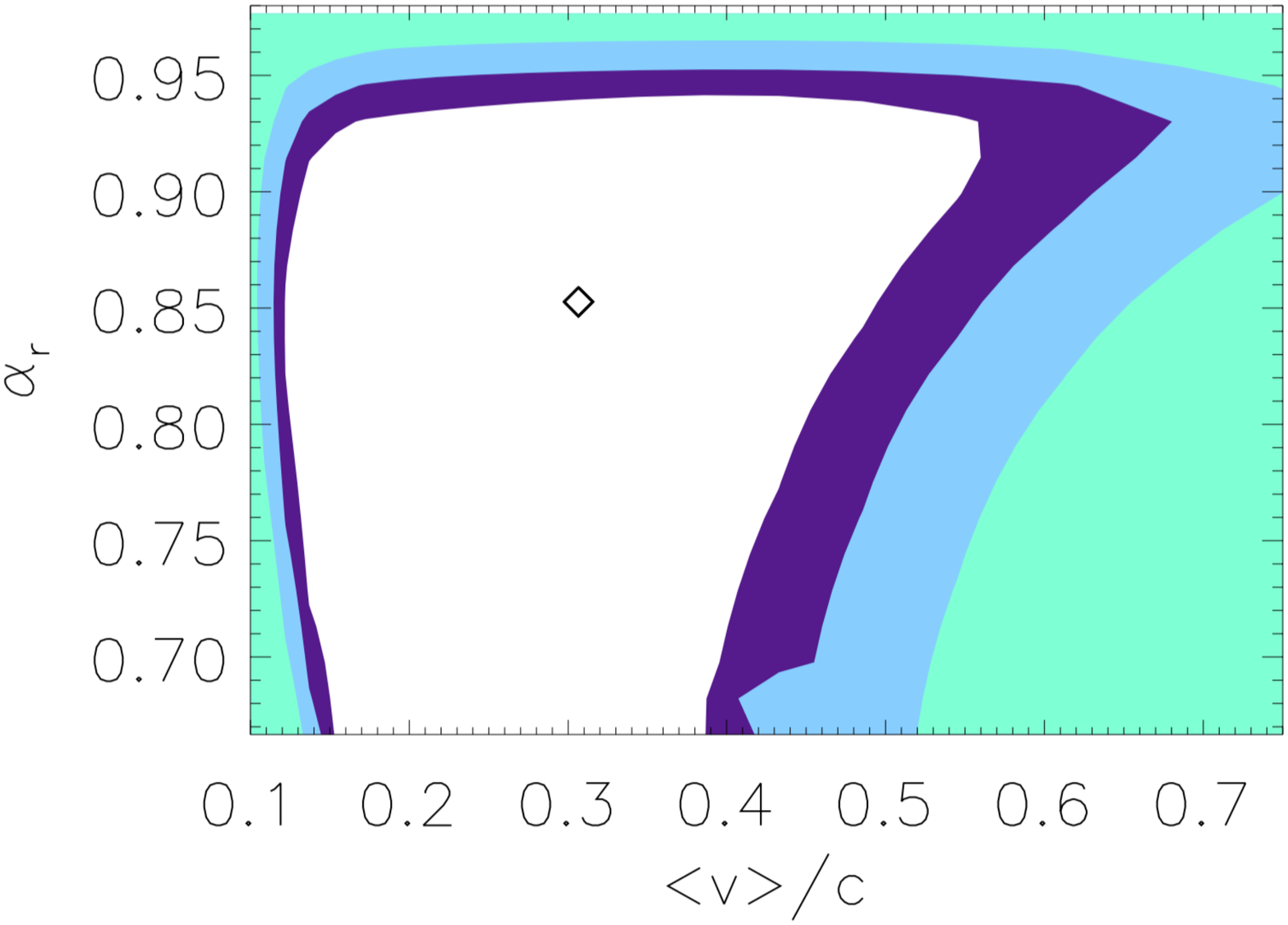}
\hspace{-0.5cm}
\includegraphics[width=7.1cm,angle=-90]{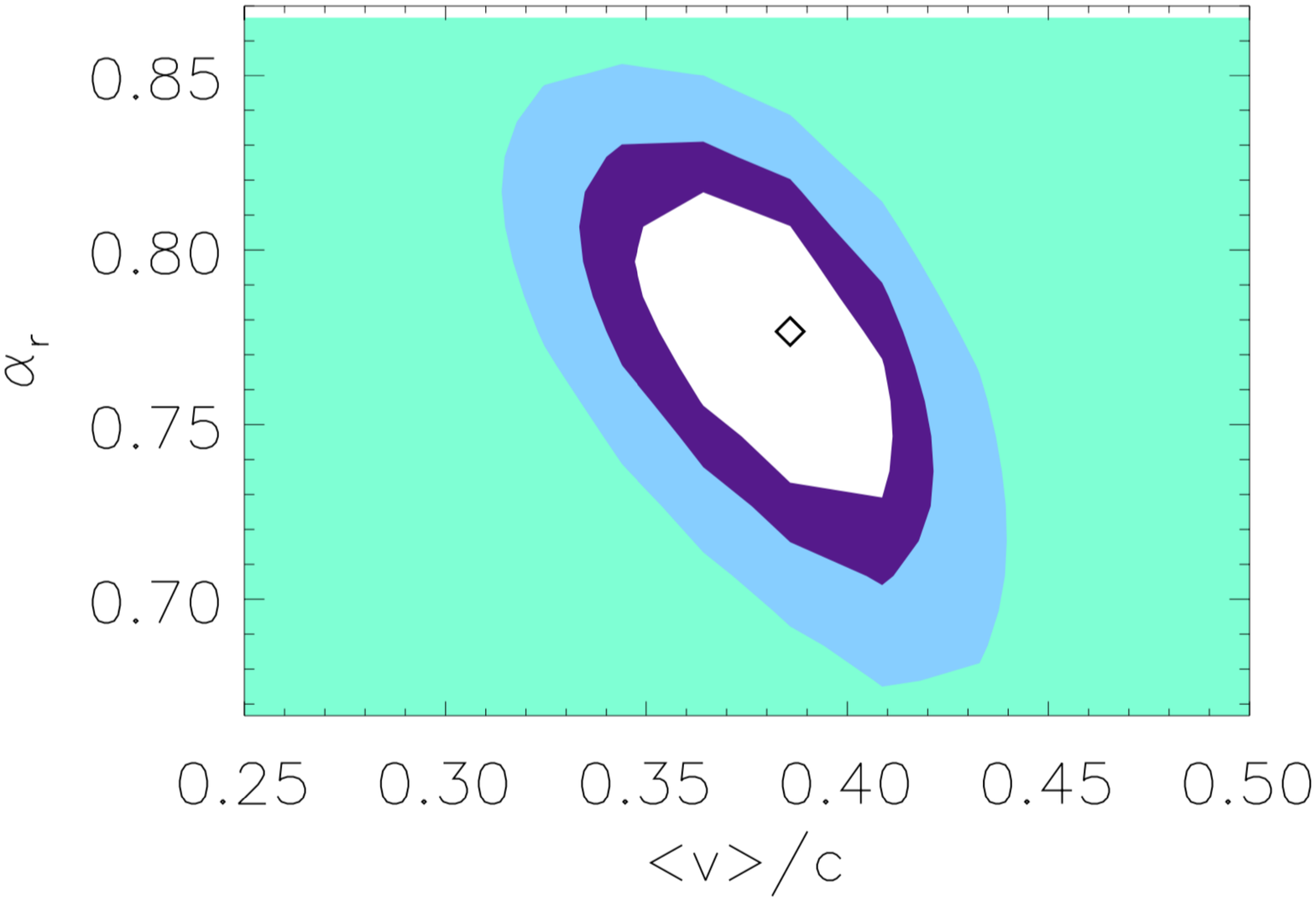}}
\caption{PTF11cmh (LEFT) and PTF14dby (RIGHT) best fit results (diamonds) and confidence intervals for the average speed and temporal index of the  blastwave radius $\alpha_r$. We expect $\alpha_r\approx 1$ for an undecelerated explosion, and $\alpha_r\approx 2/3$ for a decelerated explosion in the Sedov-Neumann-Taylor phase \citep{Waxman2004}. Colors correspond to the following confidence intervals:  $\lesssim 68\%$ confidence (white), between $68\%$ and 90\% confidence (purple), between $90\%$ and $99\%$ confidence (light blue), and  $\gtrsim 99\%$ confidence (aqua green) i.e., contours correspond to $\Delta \chi^2=2.3,4.61,9.21$ for 2 interesting parameters, respectively. The contours avoid the portions of the parameter space where the model's physical assumptions break down (i.e., the index $p$ of the electron energy distribution reaches its boundary value of $p=2$).   \label{Fig:contours}}
\end{center}
\end{figure*}

\begin{figure*}
\begin{center}
\includegraphics[width=13cm,angle=-90]{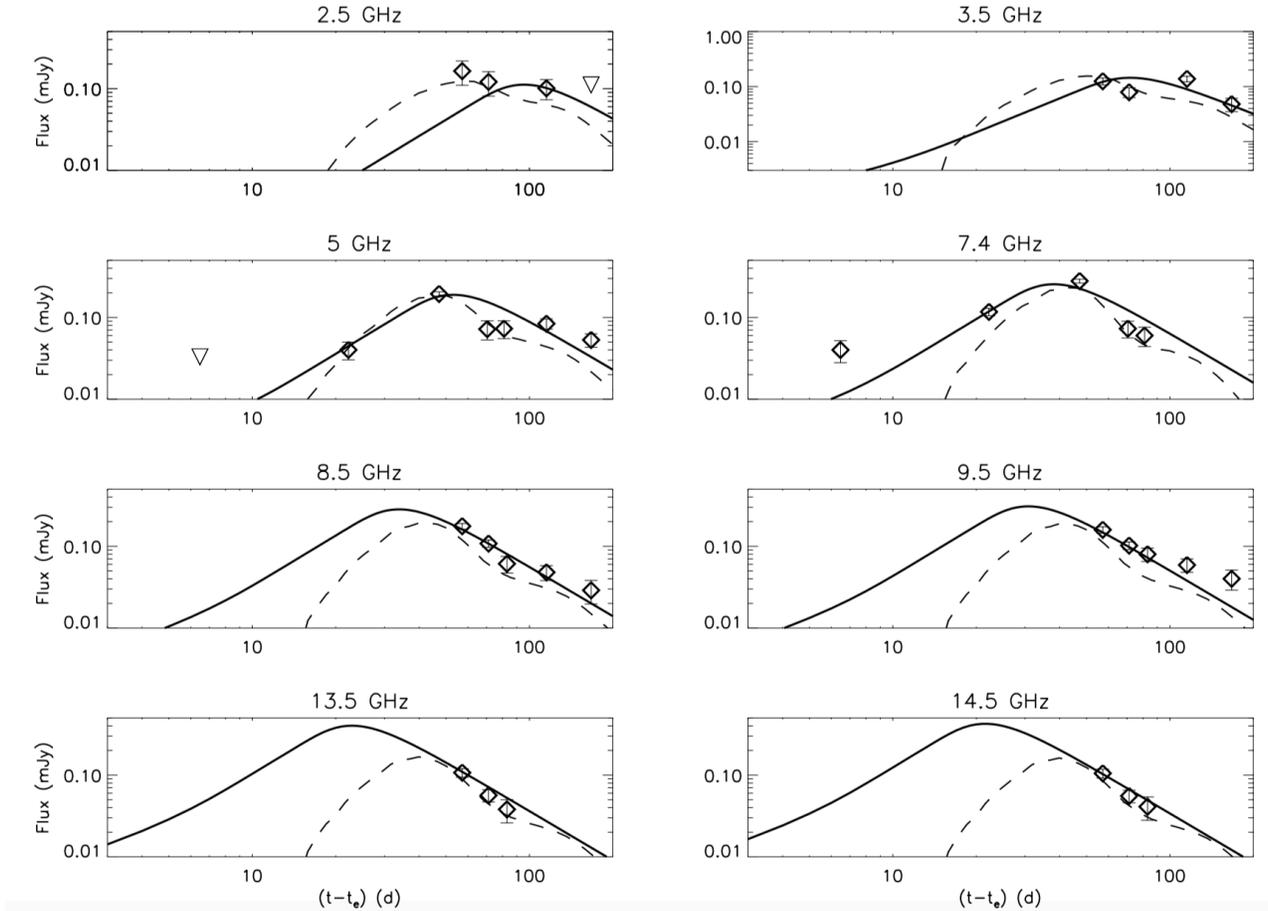}
\caption{Best fit radio light curves of PTF14dby in the synchrotron self-absorbed radio SN model (solid; see Section \ref{nonrelradio}), and an off-axis GRB model light curve for a fireball expanding in a constant density ISM (dashed; see Section \ref{vaneertenmodels}), compared with our VLA observations (see Table \ref{radioTab}; note that the two data points at $\approx 6$\,GHz are not plotted here but have been included in the fit that we use to derive the solid light curves).  See text for a discussion of the models and best fit parameters. \label{Fig:radiofits14dby}}
\end{center}
\end{figure*}

Non-thermal (self-absorbed) synchrotron radiation can be emitted from SN or GRB ejecta during  interaction with circumstellar medium (CSM). The temporal and spectral evolution of the synchrotron emission is determined by the dynamics, and by the properties of the ejecta and CSM. While young, non-relativistic SNe expand freely (their ejecta are largely un-decelerated), GRB relativistic blast waves expand and decelerate following the BM solution. At late enough times, both non-relativistic radio SNe and GRBs are expected to approach the non-relativistic adiabatic expansion phase (SNT dynamics, see also Section \ref{offaxisconstraints}). In what follows, we discuss the SNe with radio detections in our sample within these two scenarios (decelerated GRB ejecta and non-relativistic radio SN).

\subsection{GRB jets observed off-axis?}
\label{vaneertenmodels}
Three SNe in our sample, PTF11cmh, PTF11qcj, and PTF14dby, were detected during our radio follow-up with the VLA.  Interestingly, two out of these three SNe  (PTF11qcj and PTF14dby) are found to be spectroscopically similar to SN\,1998bw. Thus, the rate of radio detections for the BL-Ic SNe in our sample \textit{spectroscopically} most similar to SN\,1998bw is $\approx 2/4= 50\%$ (see Fig. \ref{BLIc:spectra}). As noticed in Section \ref{sec:XRT}, the X-ray upper-limit on PTF14dby does \textit{not} exclude the presence of X-ray afterglow emission comparable to that of GRB\,980425, however its radio emission appears different from SN\,1998bw in the fact that it peaks at later times.  Moreover, the radio emission from PTF11cmh, PTF11qcj, and PTF14dby is orders of magnitudes dimmer than that of an average long GRB observed on-axis (blue curve shaded are in Fig. \ref{earlyradio}), and also dimmer than most low-luminosity GRBs with well-sampled radio light curves. Thus, here we address the question of whether the radio emission from these three SNe could be associated with a GRB observed off-axis. 

In Figs. \ref{Fig:radiofits} and \ref{Fig:radiofits14dby} we show a tentative comparison of the observed radio light curves of PTF14dby and PTF11cmh with numerical model light curves of off-axis low-luminosity GRBs expanding in a constant density environment of density $n_{\rm ISM}$ \citep[dashed lines;][]{van2011,van2012}.   We note that numerical models for GRB jets expanding in a wind environment are not currently available to the community (at least not in a format that can allow us to easily compare these models with our observations). Thus, hereafter we limit our discussion to the case of a constant density ISM.

For PTF11cmh (Fig.  \ref{Fig:radiofits}, dashed line), we have set: $\theta_j\approx10$\, deg, observer's angle $\theta_{obs}\approx 90$\,deg, beaming corrected energy $E_{51}\approx0.1$, $n_{\rm ISM}=10$\,cm$^{-3}$, $\epsilon_B=\epsilon_e\approx 0.1$, and $p\approx 2.2$. These values for the model parameters provide a model light curve in agreement with the (limited) 6\,GHz data. We also point out that the limited dataset available for PTF11cmh does leave open the possibility of a mildly-relativistic event (discussed in more detail in the following Section).  

For PTF14dby (Fig.  \ref{Fig:radiofits14dby}, dashed line) we have set: $\theta_j\approx6$\, deg, observer's angle $\theta_{obs}\approx 70$\,deg, beaming corrected energy $E_{51}\approx0.01$, $n_{\rm ISM}=10$\,cm$^{-3}$, $\epsilon_B=\epsilon_e\approx 0.1$, and $p\approx 2.4$. While the simplest off-axis GRB model in a constant density ISM does not provide a perfect match, the model light curves are broadly compatible with the observations of PTF14dby, thus an off-axis GRB cannot be securely ruled out. We also note that in the PTF14dby radio light curve there is a hint for a late-time peak (or flattening) reminiscent of SN\,1998bw, that may be better fitted using off-axis GRB models expanding in a wind environment, and/or by invoking an energy injection episode similar to what has been proposed by \citet{Li1999} for SN\,1998bw. 

PTF11qcj is the most difficult to interpret within the simplest off-axis GRB models \citep[see also][]{Corsi2014} due to the clear late-time radio re-brightening. However, this late-time re-brightening requires modifications also to the simplest non-relativistic radio SN model, such as the presence of a denser CSM shell \citep[e.g.,][]{Salas2013}. 

Based on the tentative comparison with available models described in this Section, and on the results described in Section \ref{offaxisconstraints}, we can attempt to constrain the fraction of BL-Ic SNe in our sample potentially harboring off-axis GRB jets. Indeed, since $\lesssim 3 $ BL-Ic SNe in our sample may be associated with off-axis (low-luminosity) GRBs expanding in an ISM with $n_{\rm ISM} \sim 10$\,cm$^{-3}$ \citep[or $A_*\sim 4$; compare Eqs. (14) and (15) in][]{Waxman2004}, we set a 99.865\% confidence Poisson upper-limit of $\lesssim 12.68/15\approx 85\%$ on the fraction of BL-Ic SNe possibly harboring off-axis GRBs expanding in media with densities of this order. We note, however, that a comparison with numerical models for off-axis low-luminosity GRBs expanding in a wind environment would be needed to better determine the values of $A_*$ constrained by our dataset.

\subsection{Non-relativistic radio SN emission}
\label{nonrelradio}
In what follows, we model the radio emission observed from PTF11cmh and PTF14dby within the standard radio SN model based on the interaction of non-relativistic ejecta with CSM deposited via a constant mass-loss rate, constant velocity wind (i.e., $\rho_{\rm CSM}= \dot{M_w}/(4\pi v_wr^2)$) from a massive progenitor \citep{Chevalier1982}. We follow the formulation of this standard model given in \citet{Soderberg2003L}, which replaces the SNT dynamics with a general parametrization of the shock evolution which enables us to model the early SN synchrotron emission (when the ejecta is very close to free expansion), while recovering (in the appropriate time limit) the correct behavior for GRBs transitioning to the sub-relativistic adiabatic expansion phase \citep[][]{Waxman2004}. 

In the standard model, synchrotron emission observed at time $t$ is produced from an expanding spherical shell of shock-accelerated electrons with radius $r$ and thickness $r/\eta$. The shell interacts with a smooth CSM following a self-similar evolution. The electrons, which are accelerated into a power-law energy distribution $N(\gamma)\propto \gamma^{-p}$ (with $\gamma \gtrsim \gamma_m$), carry a fraction $\epsilon_e$ of the energy density of the ejecta. Magnetic fields carry a fraction $\epsilon_B$ of the energy density.  The temporal evolution of the shell and its properties is parametrized as \citep{Soderberg2003L,Soderberg2003bg}:
\begin{eqnarray}
\label{scaling1}r= r_0 \left(\frac{t-t_e}{t_0}\right)^{\alpha_r}~~&~~B=B_0 \left(\frac{t-t_{e}}{t_0}\right)^{\alpha_B}~~\\ 
\label{scaling2}\gamma_m=\gamma_{m,0}\left(\frac{t-t_{e}}{t_0}\right)^{\alpha_{\gamma}}~~&~~\frac{\epsilon_e}{\epsilon_B},=\mathfrak{F}_0 \left(\frac{t-t_{e}}{t_0}\right)^{\alpha_{\mathfrak{F}}}.
\end{eqnarray}
In the above Equations, $t_0$ is an arbitrary reference time (here set to day 10 since explosion), $t_e$ is the explosion time of the SN, $\alpha_r=(n-3)/(n-s)$ \citep{Chevalier1982,Chevalier1996} with $n$ the power-law index of the outer SN ejecta density profile ($\rho_{\rm SN} \propto (r/t)^{-n}$), and $s$ the power-law index of the shocked CSM electrons density profile ($n_e \propto r^{-s}$).

Following \citet{Chevalier1996}, the magnetic energy density ($U_B\propto B^2$) and the relativistic electron energy density ($U_e\propto n_e \gamma_m$) are assumed to be a fixed fraction (i.e., $\alpha_{\mathfrak{F}}=0$) of the total post-shock energy density ($U\propto n_e \left<v\right>^{2}$, where $v$ is the velocity). Making the additional conservative assumption that the energy of the radio emitting material is partitioned equally into accelerating electrons and amplifying magnetic fields ($\epsilon_e=\epsilon_B$, which implies $\mathfrak{F}_0=1$ ) and assuming $s=2$ (as expected for a wind density profile), we have \citep{Soderberg2003L,Soderberg2003bg}:
\begin{equation}
U_e\propto  U~~\Rightarrow \alpha_{\gamma}=2(\alpha_r-1),\label{ue}
\end{equation}
and 
\begin{equation}
U_B \propto U~~\Rightarrow \alpha_B=\frac{(2-s)}{2}\alpha_r-1=-1
\label{ub},
\end{equation} 
and, for the flux density from the uniform shell of radiating electrons \citep{Soderberg2003L,Soderberg2003bg}:
\begin{eqnarray}
\nonumber f_{\nu}=C_f\left(\frac{t-t_e}{t_0}\right)^{(4\alpha_r+1)/2}(1-\exp(-\tau_{\nu}))\\\times\left(\frac{\nu}{\rm 1\,GHz}\right)^{5/2}~\times F_3(x)F^{-1}_2(x)\,{\rm mJy},~~
\label{flussoteoradio}
\end{eqnarray}
where  $C_f=C_f(r_0,B_0,p)$ \citep[see Eq. (A.13) in][]{Soderberg2003L}, $x=2/3(\nu/\nu_m)$, and
\begin{eqnarray}
\nonumber \nu_m=\gamma^{2}_m \frac{e B}{2\pi m_e c}=\gamma^{2}_{m,0} \frac{e B_0}{2\pi m_e c}\left(\frac{t-t_e}{t_0}\right)^{2\alpha_{\gamma}+\alpha_B}\\
=\nu_{m,0} \left(\frac{t-t_e}{t_0}\right)^{4\alpha_{r}-5}
\label{num}
\end{eqnarray}
 is the characteristic synchrotron frequency of electrons with Lorentz factor $\gamma_m$. As typically assumed for radio SNe, we set $\nu_{m,0}\approx 1$\,GHz (which, in turn, implies that $\gamma_{m,0}$ is a function of $B_0$ only). In Equation (\ref{flussoteoradio}), $F_2$ and $F_3$ are integrals of the modified Bessel function of order 2/3 \citep[see Equation (A11) in][]{Soderberg2003L}; and
\begin{eqnarray}
\nonumber \tau_{\nu}(t)=C_{\tau}\left(\frac{t-t_{e}}{t_0}\right)^{(p-2)\alpha_{\gamma}+(3+p/2)\alpha_B+\alpha_r}\\
\nonumber \times\left(\frac{\nu}{\rm 1\,GHz}\right)^{-(p+4)/2}F_2(x)=\\
C_{\tau}\left(\frac{t-t_{e}}{t_0}\right)^{(2p-3)\alpha_r-(5p/2-1)}\left(\frac{\nu}{\rm 1\,GHz}\right)^{-(p+4)/2}F_2(x)~~~~~
\label{eq:tau}
\end{eqnarray}
is the optical depth \citep{Soderberg2003L,Soderberg2003bg}, with $C_{\tau}=C_{\tau}(r_0, B_0, \gamma_{m,0}, p)$ \citep[see Eq. (A.14) in][]{Soderberg2003L}. 
Thus, as evident from Eqs. (\ref{flussoteoradio}) and (\ref{eq:tau}), the observed spectral and temporal evolutions of the radio emission ultimately depend on the parameters $(r_0,B_{0},t_e,p,\alpha_r,\eta)$, which we determine by comparison with the data.

\subsubsection{PTF11cmh radio modeling}
Our VLA follow-up observations of PTF11cmh started at an epoch of about $\approx 20$\,d since optical discovery, and were carried out until more than $10^3$\,d after (Table \ref{radioTab}). Our first radio detection of PTF11cmh was more than 100\,d since optical discovery. 

Because our radio observations for PTF11cmh are very limited, we expect any model fitting to return only tentative estimates of model parameters. Within the standard synchrotron self-absorbed scenario (Section \ref{nonrelradio}), we can set $t_e=55673.336$\,MJD (see Table 1) and $\eta=5$ (as typically assumed in radio SN studies). This leaves 4 free model parameters to be compared with 4 radio detections and 1 upper-limit (see Table \ref{radioTab}). We thus attempt a crude fit by considering the upper-limit at epoch $\approx 20$\,d since explosion as a data point with flux value equal to the maximum radio flux detected in circular region centered on the optical position of PTF11cmh with radius equal to half the VLA FWHP synthesized beam for the observation, and error equal to the image rms i.e., $15\pm24\,\mu$Jy. This way our fit returns a $\chi^2\approx 4$ for 1 d.o.f.\footnote{We do not expect the simplified analytical synchrotron model to provide a perfect fit, and this value of the $\chi^2$ is similar to what obtained in other analyses of radio SN light curves.}. From the best fit light curves shown in Fig. \ref{Fig:radiofits} (solid lines), we also estimate $\nu_p\approx 5$\,GHz at $\approx 100$\,d since explosion, and $L_{p,5\,{\rm GHz}}\approx 10^{29}$\,erg\,s$^{-1}$\,Hz$^{-1}$. The last is comparable to the radio spectral luminosity of the GRB-associated SN\,1998bw \citep{Kulkarni1998}. 

The best-fit values for the model parameters are $p\approx 3$, $B_0\approx 5$\,G, and a blast-wave radial evolution of $R\approx 9\times 10^{15}[ (t-t_e)/10\,{\rm d}]^{0.86}$\,cm. The last implies an average ejecta speed of $R/\Delta t\approx 0.3\,c$, where $c$ is the speed of light. This is $\approx 3\times$ higher than the average speed of ordinary Ib/c SNe ($\approx 0.1c$), but smaller than relativistic events such as SN\,2009bb and SN\,1998bw. In Fig. \ref{Fig:contours} we show the uncertainties on the best values of the average ejecta speed ($\left<v\right>=r_0/10$\,d) and power-law index $\alpha_r$ of the temporal evolution of the ejecta radius, as derived by mapping the difference $\Delta\chi^2=\chi^2_{2}-\chi^2_{4}$, where $\chi^2_4$ is the best fit $\chi^2$ value returned by our 4-parameter fit to the data (see above); and $\chi^2_{2}$ is the best fit $\chi^2$ value obtained when mapping the $\alpha_r$-$r_0$ space over a grid of possible values and minimizing the $\chi^2$ over  the remaining two ``non-interesting'' parameters \citep[e.g, ][]{Avni1976}. As evident from this Figure, because of the limited dataset available for this event, the speed of the radio emitting material is very poorly constrained, with the 99\% confidence region extending in the range $\left<v\right>/c\approx 0.11-2/3$.

Assuming equipartition ($\epsilon_e=\epsilon_B=0.33$), and using Eq. (14) in \citet{Soderberg2003L}, we derive a minimum energy of $E\approx6\times 10^{48}(\epsilon_e/0.33)^{-1}[ (t-t_e)/10\,{\rm d}]^{0.58}$\,erg coupled to the fastest radio-emitting outflow. This energy is at the higher end of the range derived for other radio Ib/c SNe \citep{Margutti2014}. However, we also note that because $E\propto r^3_0\propto \left<v\right>^3$, a factor of $\approx 6$ uncertainty on $\left<v\right>$ (at 99\% confidence) implies a factor of $\approx 200$ uncertainty in the estimated ejecta energy.

Finally, the estimated progenitor mass-loss rate is $\dot{M}=10^{-4}(v_w/1000\,{\rm km\,s}^{-1})$\,M$_{\odot}$\,yr$^{-1}$, where $v_w$ is the velocity of the stellar wind and where we have assumed a nucleon-to-proton ratio of $2$ \citep[see Eq. (13) in][]{Soderberg2003L}. This mass-loss rate is higher than the typical range derived for low-luminosity GRBs (see e.g. Fig. \ref{Fig:radioULWind}), and more similar to CSM-interacting BL-Ic SNe such as PTF11qcj \citep{Corsi2014}. Based on these results, we suggest that PTF11cmh is likely a CSM-interacting event similar to PTF\,11qcj. However, this conclusion has to be taken with the caveat of being derived from a very limited dataset. Indeed, since $\dot{M}\propto r^2_0\propto \left<v\right>^2$, a factor of $\approx 6$ uncertainty on $\left<v\right>$ (at 99\% confidence) implies a factor of $\approx 40$ uncertainty in the estimated mass-loss rate.

\subsubsection{PTF11qcj radio modeling}
Our VLA follow-up observations of PTF11qcj started about two weeks after optical discovery, and were carried out until $\approx 600$\,d after \citep{Corsi2014}. As described in \citet{Corsi2014}, modeling our radio observations in the standard synchrotron self-absorbed scenario yielded best-fit values of $B_0\approx 5.7$\,G for the magnetic field, and a blast-wave radial evolution of $R\approx 1.1\times 10^{16}[ (t-t_e)/10\,{\rm d}]^{0.80}$\,cm (assuming $\eta=5$). The last implies an average ejecta speed of $R/\Delta t\approx 0.4\,c$. This is $\approx 4\times$ higher than the average speed of ordinary Ib/c SNe ($\approx 0.1c$), but smaller than relativistic events such as SN\,2009bb and SN\,1998bw.  For $\eta=5$, we also estimated a minimum energy of $E\approx1.2\times 10^{49}(\epsilon_e/0.33)^{-1}[ (t-t_e)/10\,{\rm d}]^{0.4}$\,erg coupled to the fastest radio-emitting outflow, and a progenitor mass-loss rate of $\dot{M}\approx1.5\times10^{-4}(v_w/1000\,{\rm km\,s}^{-1})$\,M$_{\odot}$\,yr$^{-1}$, where $v_w$ is the velocity of the stellar wind (and assuming a nucleon-to-proton ratio of $2$).
 
\subsubsection{PTF14dby radio modeling}
Our VLA follow-up observations of PTF14dby started at an epoch of about $\approx 6$\,d since optical discovery, and were carried out until more than $150$\,d after (Table \ref{radioTab}). The first clear ($\gtrsim 4\sigma$) VLA detection of PTF14dby at  $5$\,GHz was obtained about $20$\,d since optical discovery.  

We collected a total of 36 detections and 2 upper-limits for PTF14dby (see Table \ref{radioTab}). We model our radio detections in the standard synchrotron self-absorbed scenario (Section \ref{nonrelradio}) using a $\chi^2$ minimization procedure where we set $\eta=5$, so we are left with 5 free model parameters.  The fit returns a $\chi^2\approx 115$ for 31 d.o.f.. From the best fit light curves shown in Fig. \ref{Fig:radiofits14dby} (solid lines) we estimate $\nu_p\approx 7.4$\,GHz at $\approx 40$\,d since explosion, and a spectral peak luminosity of $L_{p,7.4\,{\rm GHz}}\approx 3\times 10^{28}$\,erg\,s$^{-1}$\,Hz$^{-1}$. The last is $\approx 4\times$ smaller than the peak radio luminosity of the GRB-associated SN\,1998bw, but comparable to the radio peak luminosity of the engine-drive SN\,2009bb \citep[Fig. \ref{radioint};][]{Soderberg2010}. 

The best-fit values for the model parameters are $t_e\approx 56832$\,MJD (which is consistent with the discovery date reported in Table 1), $p\approx 2.9$, $B_0\approx 1.6$\,G, and a blast-wave radial evolution of $R\approx 9.8\times 10^{15}[ (t-t_e)/10\,{\rm d}]^{0.78}$\,cm. The last implies an average ejecta speed of $R/\Delta t\approx 0.38\,c$. This is $\approx 4\times$ higher than the average for ordinary Ib/c SNe ($\approx 0.1 c$), but smaller than relativistic events such as SN\,2009bb and SN\,1998bw. 

In Fig. \ref{Fig:contours} we show the uncertainties on the best values of the average ejecta speed ($\left<v\right>=r_0/10$\,d) and power-law index $\alpha_r$ obtained in a way similar to what described in the previous Section. As evident from this Figure, the 99\% confidence region for the average ejecta speed is $\left<v\right>/c\approx 0.32-0.44$.

Assuming equipartition  ($\epsilon_e=\epsilon_B=0.33$), we derive a minimum energy of $E\approx8\times 10^{47}(\epsilon_e/0.33)^{-1}[ (t-t_e)/10\,{\rm d}]^{0.34}$\,erg coupled to the fastest radio-emitting outflow. 

Finally, the estimated progenitor mass-loss rate is $\dot{M}\approx 5\times10^{-6}(v_w/1000\,{\rm km\,s}^{-1})$\,M$_{\odot}$\,yr$^{-1}$ (where again we have assumed a nucleon-to-proton ratio of $2$). This mass-loss rate is in agreement with values derived for low-luminosity GRBs (see e.g. Fig. \ref{Fig:radioULWind}), and smaller that the one derived for CSM-interacting BL-Ic SNe such as PTF11qcj \citep[$\dot{M}\approx 10^{-4}(v_w/1000\,{\rm km\,s}^{-1})$\,M$_{\odot}$\,yr$^{-1}$;][]{Corsi2014}. This, together with the fact that the simplest off-axis GRB models (dashed lines in Fig. \ref{Fig:radiofits14dby}) are in broad agreement with the radio light curve of PTF14dby, calls for a more accurate numerical modeling of this SN which is beyond the scope of this paper, but that we hope will get the attention of the community.

\section{Summary and Conclusion}
\label{conclusion} 
We have presented the P48 photometry, spectral classification, and radio/X-ray follow-up observations of 15 BL-Ic SNe discovered by the PTF/iPTF. All of the SNe in our sample exclude radio afterglows typical of long duration GRBs at cosmological distances observed on-axis. Thanks to deep VLA follow-up observations, we are able to exclude the presence of 1998bw-like (or 2009bb-like) radio emission for most of the SNe in our sample. Because radio emission traces the fastest moving ejecta, we conclude that events as relativistic as, and observationally similar to, SN\,1998bw are $\lesssim 41\%$ of the BL-Ic population (99.865\% confidence). None of our upper-limits exclude radio emission similar to the radio afterglow of GRB\,060218, which faded on timescales much faster than our VLA monitoring campaign was designed to target. 

Using the X-ray upper-limits collected via our programs, we rule out the presence of off-axis GRB jets observed slightly off-axis for some of the SNe in our sample. We also constrain the energy and density parameters of (largely) off-axis GRBs potentially harbored by the SNe in our sample for which 1998bw-like radio emission was excluded. While we can rule out the presence of GRBs as energetic as GRB\,030329 observed at large off-axis angles and expanding in a constant ISM with density $n_{0,ISM}\gtrsim 0.1$, we cannot rule out the presence of off-axis GRBs expanding in a low-density wind medium, such as the one found around GRB\,130427A.

Finally, we presented the detailed radio modeling of two radio-loud BL-Ic, PTF11cmh and PTF14dby, which add to our previous radio detection of PTF11qcj. While the ejecta speed of PTF11cmh is very poorly constrained due to the limited dataset, we constrained the speed of the radio emitting material in PTF14dby to be intermediate between that of non-relativistic BL-Ic SNe, and relativistic events such as SN\,2009bb. Because we cannot securely rule out off-axis GRB models for these three events, we set an upper-limit of $\lesssim 85\%$ (99.865\% confidence) on the fraction of BL-Ic SNe in our sample that could potentially harbor a GRB observed off-axis and expanding in a medium of density $n_{\rm ISM}\sim10$\,cm$^{-3}$. This estimate could be improved by comparing our data with numerical models for off-axis GRBs expanding in a wind medium.

In summary, our results show that the VLA (thanks to its improved sensitivity) working in tandem with surveys like the iPTF, can help us clarify key open questions regarding the GRB-SN connection (such as, what fraction of purely BL-Ic SNe can host low-luminosity GRBs) and enable us to discover more events on the dividing line between ordinary BL-Ic and relativistic GRBs. Over the course of 5 years, we have greatly enlarged the sample of BL-Ic SNe (discovered independently of a GRB trigger) with radio follow-up within one year since discovery.  We expect that the Zwicky Transient Factory will be able to boost even further the rate at which we are discovering the rare BL-Ic events \citep{Smith2014}.

\bibliographystyle{apj}
\bibliography{CORS0713_r2.bib}

\acknowledgements
 A.C. thanks P. Chandra for graciously providing the data for the average long GRB radio light curve showing in Fig. 4. A.C. acknowledges support from the NSF CAREER award \#1455090. A.C. and N.P. acknowledge partial support from NASA/Swift Cycle 10 and 11 GI via grants  NNX15AB79G and NNX16AC12G. A.G.-Y.'s team is supported by the EU/FP7 via ERC grant no. 307260, the Quantum Universe I-Core program by the Israeli Committee for planning and budgeting and the ISF; by Minerva and ISF grants; by the Weizmann-UK ``making connections'' program; and by Kimmel and ARCHES awards. K.M. acknowledges support from the STFC through an Ernest Rutherford Fellowship. M.M.K. acknowledges partial support from the GROWTH project via NSF grant \#1545949. M. S. acknowledges support from the Royal Society and EU/FP7-ERC grant no. 615929. The Karl G. Jansky Very Large Array is operated by NRAO, for the NSF under cooperative agreement by Associated Universities, Inc. W. M. Keck Observatory, is operated as a scientific partnership among the California Institute of Technology, the University of California and the National Aeronautics and Space Administration. The Observatory was made possible by the generous financial support of the W. M. Keck Foundation. The William Herschel Telescope is operated on the island of La Palma by the Isaac Newton Group in the Spanish Observatorio del Roque de los Muchachos of the Instituto de Astrofísica de Canarias.   The authors acknowledge the High Performance Computing Center (HPCC) at Texas Tech University at Lubbock (\url{http://cmsdev.ttu.edu/hpcc}) for providing HPC resources that have contributed to the research results reported within this paper. This research also used resources of the National Energy Research Scientific Computing Center, a DOE Office of Science User Facility supported by the Office of Science of the U.S. Department of Energy under Contract No. DE-AC02-05CH11231.\newpage
\begin{footnotesize}
\begin{center}
\begin{longtable*}{llllllll}
\caption{VLA observations. For non-detections, the quoted UL are at $3\sigma$ (where $\sigma$ is the image rms) unless otherwise stated.  \label{radioTab}}
\\
\hline
\hline
PTF  & $T_{\rm VLA}$\tablenotemark{j}& $\Delta T_{\rm VLA}$\tablenotemark{k} & Conf. & $\nu$ & BW & Flux  & Reference \\
Name     & (MJD)           & (d)       &      & (GHz) & (MHz)  & ($\mu$Jy) & \\
\endhead
\hline
10bzf &  55268.222 & 18 & D & 5.0 & 256 & $\lesssim 33$ & \cite{ATEL2483}\\
''       &  55337.221 & 87 & D  & 6.0 & 1024 & $\lesssim 36$ & \cite{Corsi2011}\\
''       &  55527.490 & 277 & C  & 5.0 & 256 & $\lesssim 35$ & \cite{Corsi2011} \\
\hline
10qts &  55426.028 & 13 & D & 8.5 & 256 & $\lesssim 86$ & \cite{ATEL2817} \\
"        &  55948.524 & 535 & DnC & 6.2 & 2048 & $\lesssim 39$ & This paper\\
''       &  55950.435 & 537 & DnC & 6.2 & 2048 & $\lesssim 28$& This paper \\
\hline
10xem & 55767.647 & 297 &  A  & 5.0 & 256  & $\lesssim 66$ & This paper\\
\hline
10aavz &   55566.527   & 52 &  C  &    4.9 & 256     & $\lesssim 31$ & \citet{ATEL3101}\\
"            &  55770.969  & 256  & A   &   4.9 & 256   & $\lesssim 105$ & This paper\\
\hline
11cmh &  55692.145 & 19 & B  &  8.4   & 256 & $\lesssim  72$ & This paper \\
''       & 55949.356 & 276 & DnC &  6.3   & 2048 & $159\pm11$ & This paper\\
''       & 56766.099 & 1093 & A &  6.2   & 2048 & $17.4\pm5.4$ & This paper\\
"       &   56781.411   &   1108    & "  &  1.7  & "       & $57\pm13$ & This paper\\
"      &  56808.981 &   1136   & "    &     2.9    & "       &     $18.0\pm4.2$  & This paper\\
\hline
11img &  55781.981 & 31 & A & 5.0   & 256 & $\lesssim 48$ & This paper\\
''       &  55950.478 & 199 & DnC &  5.0  & 2048 & $\lesssim 66$& This paper\\
\hline
11lbm & 55812.283 &  19  &  A & 5.0  & 256  & $\lesssim 78$ & This paper\\
''       &  55948.043 & 155 & DnC & 6.2   & 2048 & $\lesssim28$ & This paper\\
\hline
12as &  55933.555 & 8 & DnC & 6.2  & 2048 & $\lesssim 87$ &This paper \\
''      & 55947.382 & 22 & DnC &  6.2 & 2048 & $\lesssim 69$ & This paper\\ 
''      & 56999.567 & 1074 & C &  6.2 & 2048 & $\lesssim 75\footnote{$5\sigma$ UL corresponding to the brightness of the host galaxy.} $ & This paper\\ 
\hline
13u &  56391.376 & 67 & D &  6.2 & 2048 & $\lesssim 95$\footnote{$10\sigma$ UL corresponding to the brightness of the host galaxy.} & This paper\\
''     & 56767.366 & 443 & A & 6.2 & 2048 & $\lesssim 14$ & This paper \\
''     &  57311.028&  987 & D & 6.2 & 2048 & $\lesssim 100$\footnote{$10\sigma$ UL corresponding to the brightness of the host galaxy.}  & This paper \\
\hline
13alq & 56401.120 & 7 & D& 4.8  & 2048 & $\lesssim 30$& \cite{ATEL4997}\\
13alq &  56423.080 & 29 & DnC & 6.2  & 2048 & $\lesssim 60$\footnote{3.6$\sigma$ UL.} & This paper\\
13alq &   57336.833 &  942 & D & 6.3 & 2048 & $\lesssim 33$ & This paper\\
\hline
13ebw &  56667.566   & 46 & B & 6.2 & 2048 & $\lesssim 15$ & This paper\\
" &    57001.417   &  380   & C &    6.2 &  2048 &     $\lesssim 41$\footnote{7$\sigma$ UL corresponding to the brightness of the host galaxy.}                & This paper\\
\hline
14dby &    56838.197  &   6 & D & 5.2 & 1024 & $\lesssim 33$ & This paper\\
"                 &   "                  &   "  & " & 7.5 & " & $40\pm12$ & " \\
"                &   56853.994  & 22  & " & 5.2 & " & $40.1\pm9.8$ & "\\
"                &   "                  & "    & " & 7.5 & " & $117\pm12$ & "\\
"                &  56879.117    & 47  & " & 5.0 & " & $194\pm12$ & "\\
"                &  "                    & "     & " & 7.4 & " & $279\pm15$ & "\\
"                &  56889.108    & 57  & " & 2.5 & " & $164\pm54$ & "\\
"                &  "                    & "    & " & 3.3 & " & $125\pm18$ & "\\
"               &  "                    & "    & " & 8.5 & " & $176\pm14$ & "\\
"                &  "                    & "    & " & 9.5 & " & $159\pm14$ & "\\
"                &  "                    & "    & " & 13.5 & " & $107\pm13$ & "\\
"                &  "                    & "    & " & 14.5 & " & $105\pm13$ & "\\
"                &  56902.138    & 70  & " & 5.0 & " & $72\pm19$ & "\\
"                &  "                    & "     & " & 7.4 & " & $73\pm17$ & "\\
"                &  56903.045    & 71   & " & 2.7 & " & $121\pm40$ & "\\
"                &  "                    & "     & " & 3.2 & " & $79\pm16$ & "\\
"                &  "                    & "     & " & 8.5 & " & $108\pm11$ & "\\
"                 &  "                    & "     & " & 9.5 & " & $102\pm11$ & "\\
"                 &  "                    & "     & " & 13.5 & " & $56.5\pm9.6$ & "\\
"                 &  "                    & "     & " & 14.5 & " & $55.5\pm9.7$ & "\\
"                &  56913.144    & 81   & " & 5.1 & " & $73\pm18$ & "\\
"                &  "                    & "     & "  & 7.5 & " & $60\pm16$ & "\\
"                &  56914.770     & 83  & "  & 8.5 & " & $61\pm14$ & "\\
"                &  "                    & "     & "  & 9.5 & " & $80\pm15$ & "\\
"                &  "                    & "     & "  & 13.5 & " & $38\pm12$ & "\\
"                &  "                    & "     & "  & 14.5 & " & $41\pm13$ & "\\
"                &  56946.949     & 115  & DnC  & 2.5 & " & $101\pm28$ & "\\
"                &  "                     & "  & "  & 3.5 & " & $138\pm16$ & "\\
"                &  "                     & "  & "  & 5.0 & " & $84\pm12$ & "\\
"                &  "                     & "  & "  & 6.0 & " & $79\pm15$ & "\\
"                &  "                     & "  & "  & 8.5 & " & $48\pm10$ & "\\
"               &  "                     & "  & "  & 9.5 & " & $59\pm11$ & "\\
"                &  56998.800      & 167 & C  & 2.5 & " & $\lesssim 111$ & "\\
"                &  "                     & "  & "  & 3.4 & " & $48\pm13$ & "\\
"                &  "                     & "  & "  & 5.0 & " & $53\pm10$ & "\\
"                &  "                     & "  & "  & 6.0 & " & $44\pm14$ & "\\
"                &  "                     & "  & "  & 8.5 & " & $28.9\pm9.3$ & "\\
"                 &  "                     & "  & "  & 9.5 & " & $40\pm11$ & "\\
\hline
PTF14gaq &  56932.496   &  8 & DnC &   6.3  & 2048 & $\lesssim 28$ & This paper\\
"               &  56998.857    &   75& C &    6.2 & 2048 & $\lesssim 25$ & This paper\\
"                & 57335.919    &      411  & D &   6.3 & 2048 & $\lesssim 19$ & This paper\\
\hline
PTF15dld &   57336.049 & 18 &   D &  5.0& 2048 & $\lesssim 110\footnote{10$\sigma$ UL corresponding to the brightness of the host galaxy.}$ & This paper\\
\hline
\multicolumn{8}{l}{$^j$ The VLA observation time $T_{\rm VLA}$ is the time at the mid-point of the VLA observation.}\\
\multicolumn{8}{l}{$^k$ VLA observation epoch in days since PTF discovery, not corrected for redshift effects.}
\end{longtable*}
\end{center}
\end{footnotesize}

\end{document}